\documentclass[10pt, a4paper]{article}
\usepackage {fullpage}						
\usepackage {ucs}
\usepackage [utf8x]{inputenc}
\usepackage {graphicx}

\usepackage {amsfonts,amsmath,amsthm,amssymb}
\usepackage {latexsym, bbm}
\usepackage {graphics, graphicx}
\usepackage {fancyhdr}
\usepackage[table]{xcolor}
\usepackage {epstopdf} 						

\usepackage {multicol}						
\usepackage {multirow}  					
\usepackage {soul} 							

\usepackage {authblk} 						
\usepackage {datetime} 						
\usepackage {subfigure}						
\usepackage {parskip}						
\usepackage {setspace}						
\usepackage[font=small,textfont=it,width=0.97\textwidth]{caption}		
\usepackage {mathtools}						
\usepackage {array}							

\usepackage{filemod}						
\usepackage{tabularx}

\renewcommand{\(}{\left(}
\renewcommand{\)}{\right)}

\newcommand{\go}{[g]_0}
\newcommand{\gb}{[g]_b}
\newcommand{\gf}{[g]_f}

\newcommand{\ro}{[r]_0}
\newcommand{\rb}{[r]_b}
\newcommand{\rf}{[r]_f}

\newcommand{\cc}{c^\text{S}}
\newcommand{\cd}{c^\text{D}}
\newcommand{\cf}{c^\text{F}}

\newcommand{\lc}{\lambda^\text{S}}
\newcommand{\ld}{\lambda^\text{D}}

\newcommand{\cs}{^\text{S, sim}}

\newcommand{\cm}{\text{cm}}

\newcommand{\seconds}{\text{s}}
\newcommand{\mol}{\text{mol}}

\newcommand{\memt}{\mu_\text{EMT}}

\newcommand{\mtra}{\mu_\text{TRA}}

\newcommand{\rfr}{^\text{ref}}

\newcommand{\sign}{\operatorname{sign}}
\newcommand{\pd}{\partial}

\def\gray#1{{\color{gray}\text{#1}}}

\def\comm#1{{\color{red}((\ul{#1}))}}

\newtheorem{experiment}{Experiment}

\title{A multiscale approach to the migration of cancer stem cells: mathematical modelling and simulations}

\author{
	Nikolaos Sfakianakis\thanks{Institute of Mathematics, Johannes Gutenberg-University, Mainz, Germany\hfill {\tt sfakiana@uni-mainz.de}}~{},
	Niklas Kolbe\thanks{Institute of Mathematics, Johannes Gutenberg-University, Mainz, Germany\hfill {\tt kolbe@uni-mainz.de}}~{},
	Nadja Hellmann\thanks{Institute of Molecular Biophysics, Johannes Gutenberg-University, Mainz, Germany\hfill {\tt nhellman@uni-mainz.de}}~{},
	M\'aria Luk\'a\v{c}ov\'a-Medvid'ov\'a\thanks{Institute of Mathematics, Johannes Gutenberg-University, Mainz, Germany\hfill {\tt lukacova@uni-mainz.de}}~{},
	}

\begin{document}
\maketitle


\begin{abstract}
	We propose a multiscale model of the invasion of the \textit{extracellular matrix} by two types of cancer cells, the \textit{differentiated cancer cells} and the \textit{cancer stem cells}. We assume that the \textit{epithelial mesenchymal-like transition} between them is driven primarily by the \textit{epidermal growth factors}. We moreover take into account the \textit{transidifferentiation} program of the cancer stem cells and the \textit{cancer associated fibroblast cells} as well as the fibroblast-driven \textit{remodelling} of the extracellular matrix. 
	
	The proposed \textit{haptotaxis} model combines the \textit{macroscopic} phenomenon of the invasion of the extracellular matrix with the \textit{microscopic} dynamics of the epidermal growth factors. We analyse our model in a component-wise manner and compare our findings with the literature. We investigate pathological situations regarding the epidermal growth factors and accordingly propose ``mathematical-treatment'' scenarios to control the aggressiveness of the tumour.  
\end{abstract}

\section{Introduction}
	The mathematical study of cancer processes has been an active field of research since the 1950s; see, e.g., \cite{Nordling.1953, Armitage.1954, Fisher.1958}. It covers a wide range of biological/bio-chemical processes of the tumour from the perspective of modelling, analysis and numerical simulations, see, e.g., \cite{Preziosi.2003,Perumpanani.1996, Anderson.2000, Gerisch.2008, Ganguly.2006,Michor.2008, Czochra.2012,7,Johnston.2010,Vainstein.2012, Gupta.2009}.
		
	The nature of the cancer cells has also been under scientific scrutiny. In particular, although cancer cells exhibit higher proliferation rates than normal cells, recent evidence has revealed the existence of a subpopulation of the tumour that proliferates with lower rates, exhibits stem-like properties such as self-renewal and cell differentiation, and is able to metastasize, i.e. to detach from the primary tumour, invade the vascular system, and afflict secondary sites \cite{Brabletz.2005, Mani.2008}. These cancer cells, termed \textit{cancer stem cells} (CSCs), constitute only a minor part of the tumour, while its bulk is comprised by the more proliferative and non-metastatic \textit{differentiated cancer cells} (DCCs), \cite{Thiery.2002, Reya.2001}. The CSCs were first demonstrated in human acute myeloid leukemia and several solid tumours, such as breast, brain, melanoma, prostate, ovarian, and pancreatic, see \cite{Fan.2013} and the references therein.

	It is not completely known, where the CSCs originate from; a current theory states that they emanate from the more usual DCCs after a \textit{de-differentiation} program has taken place \cite{Gupta.2009, Reya.2001}. This type of transition of cancer cells, influences their \textit{cellular potency} and seems to be related to the \textit{epithelial-mesenchymal transition} (EMT), a type of cellular de-differentiation program that can also be found in normal tissue, \cite{Mani.2008}. After EMT the newly sprout CSCs take the first step in metastasis and invade the \textit{Extracellular Matrix} (ECM), what is considered the one of the ``hallmarks of cancer'', cf. \cite{Thiery.2002,Katsumo.2013}.

	The trigger of EMT is set by components of the micro-environment of the tumour \cite{Radisky.2005}. A prime role in this mechanism is played by the \textit{epidermal growth factor} (EGF) that stimulates cell growth, proliferation, and differentiation by binding to the corresponding cellular receptor EGFR  \cite{3,Shien}. 
	
	This process can be briefly described as follows: the binding of an EGF molecule onto an EGFR stimulates ligand-induced dimerization that activates the protein-tyrosine kinase activity of the receptor which in turn, initiates a signal cascade that results in a variety of changes within the cell and leading to DNA synthesis. In certain pathological mutations they get stuck in the ``on'' position and cause unregulated cell growth, untimely EMT, and over-expression of EGFRs. Over-expressed EGFRs are {found} in many types of tumors, including breast, thyroid, ovarian, colon, head and neck, and brain. In addition, EGFR over-expression has been linked to a poor prognosis in breast cancer and may promote proliferation, migration, invasion, and cell survival as well as inhibition of cell apoptosis \cite{2}. The role of EGFR as oncogene \cite{Kawamoto.1983} has led to the development of anticancer therapies targeted against the EGFR, the so-called \textit{EGFR inhibitors}, that promise {improved} treatment for many types of solid tumors.

	\begin{figure}[t]
		\centering
		{\includegraphics[width=0.7\linewidth]{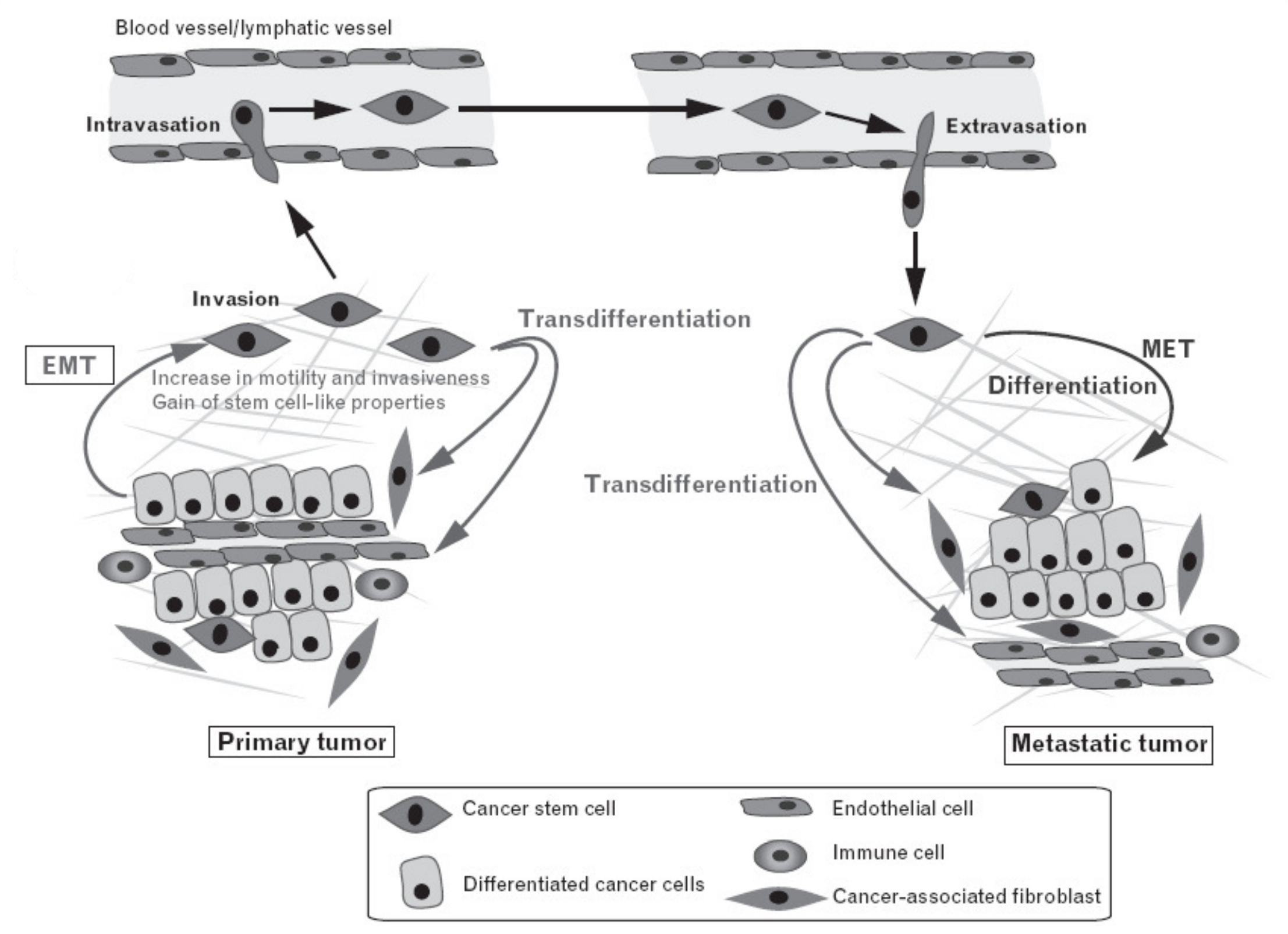}}
		\caption{Schematic representation of the CSCs' participation in the metastasis of cancer. Showing here the invasion, intravasation, and translocation of CSCs to the sites of metastasis. Showing moreover the EMT transition from DCCs to CSCs (left part of the graph), and the transdifferentiation of CSCs to fibroblasts and endothelial cells; Source  \cite{Katsumo.2013}.\label{fig:metastasis} \comm{Permission of the journal pending}}
	\end{figure}

	In this work we consider a special type of taxis, namely the \textit{haptotaxis}, in order to model the invasion of the ECM by the DCCs and CSCs. Haptotaxis describes the biased random motion along directions of either extracellular adhesion sites, or of the ECM-bound chemoattractants, or along their gradients. Haptotaxis is typically modelled under the paradigm of the Keller-Segel (KS) systems\footnote{First introduced in \cite{KS.1970} to describe the aggregations of slime mold.}. Since their initial derivation, KS systems have been successfully applied in a wide range of biological phenomena spanning from bacterial aggregations to wound healing and cancer modeling. In cancer growth modelling in particular, the Keller-Segel type systems were extended to include also enzymatic interactions/reactions, yielding \textit{advection-reaction-diffusion} (ARD) models, see for instance the generic haptotaxis system proposed in \cite{Anderson.2000} that takes into account ECM degradation by \textit{matrix metalloproteinases} (MMPs), or the chemo-hapto-taxis system proposed in \cite{Chaplain.2005} that addresses the role of the enzyme \textit{plasmin}. 

	Besides degradation, the ECM is also dynamically remodelled; a process that is essential not only for the developmental progress of healthy tissue (wound healing, organ homeostasis, etc.), but is also crucial in the metastatic progression of cancer \cite{Cox.2011}. Specifically, the degradation of the ``normal ECM'' and its replacement by ``tumour induced ECM'' leads to altered physiological conditions more advantageous for the tumour, \cite{Kaplan.2005, Erler.2009}. Main characteristic in this process is the increased expression of \textit{fibroblast cells} which result from a \textit{transdifferentiation} program of CSCs, see Figure \ref{fig:metastasis} for a graphical depiction of the aforementioned processes and refer to \cite{Katsumo.2013} for more details.
	
	Our main objective is to model the EMT transition of DCCs to CSCs, triggered by the interaction with the EGF molecules found in the extracellular environment and the cellular EGF receptors (EGFRs). In more detail, we couple the microscopic dynamics of the EGF attachment/detachment to EGFRs with the macroscopic dynamics of the ECM invasion. We assume that these microscopic processes are much faster than the kinematic properties of the cells, including their invasion of the ECM. The overall process is limited by the availability of free EGF molecules and by the number of free EGFRs on the surfaces of the cells.
	
	The ECM invasion model that we have developed, is an ARD haptotaxis system that includes both the DCCs and the CSCs, their kinematic properties, and the EMT. It moreover includes the ECM on which the cancer cells adhere. We assume that the ECM is degraded by a class of matrix degenerating \emph{metalloproteinases} (MMPs) ---secreted by the cancer cells. The ECM is also constantly remodelled by \textit{fibroblast} cells; which are assumed to emanate from the CSCs via a \textit{transdifferentiation} program \cite{Kaplan.2005, Erler.2009, Katsumo.2013}.
	
	We study the dynamics of the derived model on one- and two-dimensional computational domains. The numerical treatment of such problems is challenging due to the inherent heterogeneous spatio-temporal dynamics and the merging/emerging clusters of cancer cells that the solutions develop see, e.g., Figure \ref{fig:invasion1D}. Thus, we employ a second order finite volume method equipped with a third order time integration scheme, developed particularly for the needs of this type of problems, see Section \ref{sec:numerics} and our previous works \cite{Sfakianakis.2014b,Kolbe.2013} for more details. In our investigation we also include a sensitivity analysis of the model with respect to its parameters. In particular, we investigate the effect that the parameters have on the mass of the tumour and on the invasiveness of the DCCs and CSCs.
	
	Similar ECM invasion systems that account for non-homogeneous cancer cell populations and the transition between them, exist in the literature see, e.g., \cite{Andasari.2011,Domschke.2014, Stinner.2015}. However, in these works a detailed investigation of the EMT triggering mechanism is not included, rather it is assumed that EMT takes place with a constant ad hoc rate. With our model we can see the impact of pathological situations (such as the EGF depletion), and to identify significant differences in the production of CSCs between small and large tumours.

	The rest of the paper is structured as follows: in Section \ref{sec:model} we derive our mathematical model. We do so in a subsystem-by-subsystem way, and justify our choices in detail. In Section \ref{sec:numerics} we describe the numerical method we use to resolve our model, and the method employed in the parameter sensitivity analysis. In Section \ref{sec:discussion} we present and discuss a series of experiments aimed to highlight the influence of the various model components, to compare with existing results from the literature, and to deduce specific ``mathematical-treatment'' scenarios to control the aggressiveness of the CSCs.
	
\section{Model}\label{sec:model}
	Our mathematical model is deterministic and phenomenological. The unknown variables represent densities of the corresponding model components. More precisely, we include the two types of cancer cells (DCCs and CSCs), the ECM, the fibroblast cells that are responsible for the remodelling of the matrix, as well as matrix degenerating proteins. It consists of three subsystems a) the ECM invasion subsystem in which the basic components of the haptotaxis movement of the cancer and fibroblast cells are included, b) the EMT subsystem that is responsible for the EGF-driven transition of the CSCs to the DCCs, and c) the ECM remodelling subsystem which addresses way the fibroblast cells reconstruct the ECM. In the paragraphs that follow we present and justify each system separately, and at the end of this section we concatenate them to derive the complete system.
	 	
\subsection{The cancer invasion subsystem}
	This part of the model is based on the original derivation of a chemotaxis systems by Keller and Segel (KS), \cite{KS.1970}. Although these systems where initially developed to address the phenomenon of bacterial aggregation, they have been very successful in applications to a wide range of biological phenomena, see, e.g., \cite{Lukacova.2012} and the references therein. In modelling cancer growth, KS-type systems have been adjusted to address haptotaxis and a large variety of interactions between the cancer cells and the extracellular environment, see, e.g., \cite{Alt.1985, Bellomo.2008, Anderson.2000}. 
	
	For our model we first employ the basic haptotactic system which features a) the \emph{cellular} diffusion and logistic  proliferation of the cancer cells, b) the proteolysis of the ECM by matrix degenerating \emph{metalloproteinases} (MMPs), and c) the \textit{chemical} degradation, diffusion of the MMPs as well as their production by the cancer cells:
	
	\begin{equation}\label{eq:sub_basic}
	\left\{
	\begin{aligned}
		\frac{\pd c}{\pd t} &= D_c \Delta c -\chi\nabla \cdot\(c \nabla v\) + \mu c (1-c-v)^+\\
		\frac{\pd v}{\pd t} &=-\delta m v \\ 
		\frac{\pd m}{\pd t} &= D_m\Delta m + \alpha c -\beta m
	\end{aligned}\,,\right.
	\end{equation}
	where the \emph{positive part} function is defined as $(x)^+=\max\left\{x,0\right\}$. Here we denote the densities of the cancer cells, ECM, and MMPs by $c$, $v$, and $m$ respectively. 
	
	A first assumption we make in the model is that the cancer cells are ``immortal''; we exclude hence any terms that describe \emph{age-dependent} cell death. In a similar way we exclude from the model cell death due to \textit{overcrowding} and the competition for free space, so in effect we consider the \emph{positive-part} logistic proliferation as opposed to the more usual logistic proliferation. Note moreover that the system \eqref{eq:sub_basic}, and the ones that follow, are presented in their scaled non-dimensional version. We provide information on the scaling coefficients and reference densities along with the parameter values in Table \ref{tbl:params}. 
	
\subsection{The EMT subsystem}\label{sec:EMT.mod}

	Both species of cancer cells invade the ECM in a similar way. Indeed, we assume that they both satisfy corresponding KS-type systems of the form \eqref{eq:sub_basic}. They are also coupled by the EMT, which works as a loss/gain term between them:	
	\begin{equation}\label{eq:sub_emt}
	\left\{
	\begin{aligned}
		\frac{\pd \cd}{\pd t}&=\gray{diffusion}+\gray{haptotaxis} - \memt \;\cd
														+\gray{proliferation}\\												
		\frac{\pd \cc}{\pd t}&=\gray{diffusion}+\gray{haptotaxis} + \memt \;\cd
												+\gray{proliferation}
	\end{aligned}\right.\, ,\end{equation}
	where by $\cd$, $\cc$, and $\memt$ we denote the DCCs, CSCs, and EMT rate, respectively. As is typically done in the modelling of reaction between two components, we assume that the EMT depends on the amount of EGF ---in particular on the occupied DCC-specific EGFRs--- and on the local density of the DCCs. We assume moreover that the $\memt$ is an increasing function of the occupied EGFRs that stagnates asymptotically to a constant value i.e.:
	\begin{equation}\label{eq:muemt}
		\memt=\mu_0 \frac{\gb^\text{DCC}}{\mu_{1/2} + \gb^\text{DCC}}\,,
	\end{equation}
	where $\gb^\text{DCC}$ is the density of the EGF molecules that are bound onto the EGFRs, $\mu_0$ is the maximum EMT rate, and $\mu_{1/2}$ a is the critical amount of bound EGF $\gb^\text{DCC}$ needed to generate half of the maximum EMT rate $\mu_0$. In effect, $\mu_{1/2}$ becomes the critical measure of EGF that characterizes the non-uniform EMT. 
	
	This modelling approach is quite general as it includes both the EGF and the DCCs densities. It has been motivated by similar previous approaches, i.e. \cite{Eladdadi.2008, Zhu.2011}, and was also employed by us in \cite{Sfakianakis.2015}. Although a simplification of the biological reality, this choice of $\memt$ suffices to exhibit the importance of considering non-uniform EGF-driven EMT rate, see also Section \ref{sec:discussion}.
	
	In the rest of this section we briefly describe the modelling/computation of $\memt$, and refer to \cite{Sfakianakis.2015} for more details. We set  $\gf$ to represent the amount of EGF that is free in the extracellular environment, $\go$ the total (free and bound on both the DCCs and the CSCs) density of EGF. Accordingly, $\ro$, $\rb$, and $\rf$ denote the total, occupied, and free receptors on the surface of cancer cells. We assume that the EGF and EGFR are locally conserved:
	\begin{align}
		\go(t,x)&=\gf(t,x) + \gb(t,x)\,, \label{eq:mass.EGF}\\
		\ro(t,x)&=\rf(t,x) + \rb(t,x)\,, \label{eq:mass.EGFR}
	\end{align}
	for every $t\geq0$ and $x\in \Omega\subset\mathbb{R}$ bounded. The binding/release of EGF to/from the EGFR receptors takes place with rates $k_+$/$k_-$ and is the same for both the DCCs and the CSCs,  see also \cite{Zhu.2011}. Moreover we assume that the free EGF molecules diffuse in the extracellular environment:
	\begin{equation}\label{eq:parab}	
		\left\{ \begin{aligned}
           \partial_\tau \gb &= && k_+ \gf \rf - k_- \gb \\
           \partial_\tau \gf &= D_f \Delta \gf - && k_+ \gf \rf  + k_- \gb
          \end{aligned}\,,
		\right.
	\end{equation}
	The dynamics in \eqref{eq:parab} take place during a microscopic time scale $\tau$, which is related to the macroscopic time $t$ through:
	\begin{equation}\label{eq:time.rel}
		\tau(t) =  \frac{t}{\varepsilon}, \quad 0<\varepsilon <<1\,.
	\end{equation}

	We express by
	$\begin{pmatrix}g_b (t,x), & g_f(t,x) \end{pmatrix} = \begin{pmatrix}\gb(\tau(t),x), & \gf(\tau(t),x) \end{pmatrix}$ 
	the variables of the parabolic system \eqref{eq:parab} in the $t$ timescale
	and recall \eqref{eq:mass.EGFR}, to write the amount of free receptors as
	$$\rf\ =\ \ro - \rb\ =\ \lc \cc + \ld \cd - \gb,$$ 
	where $\lc, \ld$ represent the total (free and bound) EGFR receptors per cell. We have assumed that each EGFR receptor can be occupied by a single EGF molecule, i.e. $\rb=\gb$.  We can now deduce from \eqref{eq:parab}, after taking the formal limit as $\varepsilon\rightarrow 0$ in \eqref{eq:time.rel}, the elliptic system:
	\begin{equation}\label{eq:ellip}
		\left\{ \begin{aligned}
			\frac{1}{k_D}\, g_f\, \(\lc\cc +\ld\cd - g_b\) - g_b = 0\\
			\Delta g_f = 0
		\end{aligned} \,, \right.
	\end{equation}
	where $k_D={k_-}/{k_+}$, the attachment/detachment ratio of the EGFs onto the EGFRs. The system \eqref{eq:ellip} is also augmented with homogeneous Neumann boundary conditions, 
	\begin{equation}\label{eq:neumann}
		\frac{\pd g_f(x)}{\pd \bf n} = 0,\quad x\in\pd \Omega\,.
	\end{equation}
	where $\bf n$ is the outward unit normal vector to the computational domain of the equation \eqref{eq:ellip}. The uniqueness of the solution of \eqref{eq:ellip}-\eqref{eq:neumann} is recovered by the additional assumption that the total amount of EGF remains constant,
	\begin{equation}\label{eq:egf_mass}
		\Gamma = \frac{1}{ |\Omega|}\int_\Omega g_0(t,x)\,dx= \frac{1}{ |\Omega|} \left( \int_\Omega g_b(t,x)\,dx + \int_\Omega g_f(t,x)\, dx \right)\,.
	\end{equation}
	
\begin{figure}[t]
		\centering
		\resizebox{\linewidth}{!}{
		\includegraphics{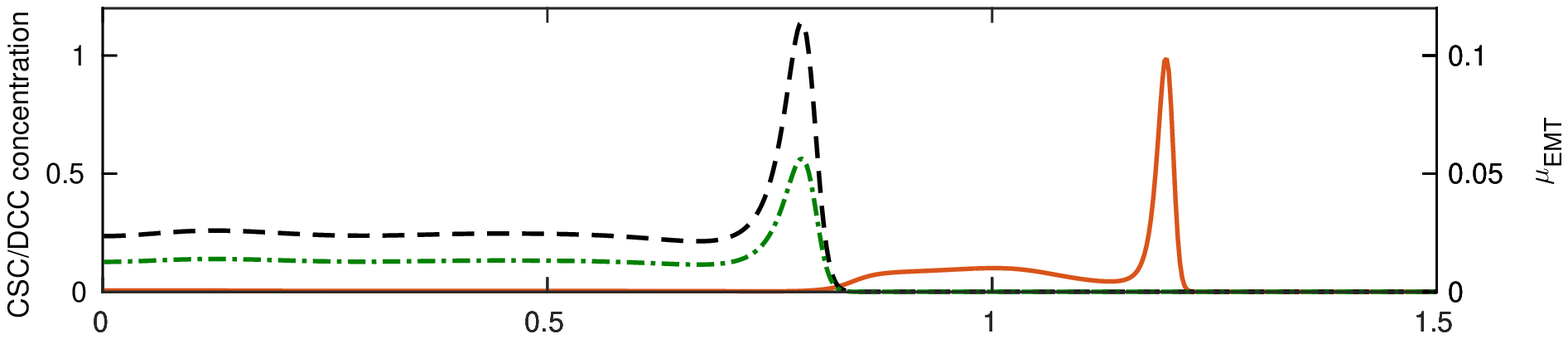}}
	\vfill \vspace{.25em}
	\includegraphics[scale=0.8, trim= 0 -2.5pt 0 0]{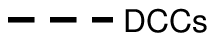}\quad
	\includegraphics[scale=0.8, trim= 0 -2.5pt 0 0]{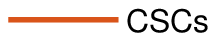}\quad
		\includegraphics[scale=0.8]{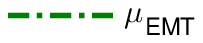}
		\caption{The coefficient $\memt$ depends non-uniformly on the space variable (here one-dimensional). This follows from the dependence of $\memt$ on $\cc$ \eqref{eq:muemt_2} which makes it also time-dependent, for more details see \cite{Sfakianakis.2015}.}\label{fig:memt}
		
	\end{figure}

	With the above assumptions the system \eqref{eq:ellip} can be solved to give
	\begin{equation}\label{eq:g_b_full}
		g_b = \frac{g_f}{ k_D + g_f} \(\lc\cc +\ld\cd\)\,,
	\end{equation}
	and in particular
	\begin{equation}\label{eq:g_b}
		g_b^\text{DCC}  = \frac{g_f}{k_D + g_f} \ld\cd\,.
	\end{equation}
	When combined with \eqref{eq:muemt} we obtain
	\begin{equation}\label{eq:muemt_2}
		\memt=\mu_0\frac{g_f\ld\cd}{\mu_{1/2}k_D + \mu_{1/2} g_f + g_f\ld\cd}\,,
	\end{equation}
	which is non-uniform in space and time, see also Figure \ref{fig:memt}. On the other hand, the free EGF $g_f$ is constant in space and is given by the unique positive root of the quadratic equation (see also \cite{Sfakianakis.2015} for more details):
	\begin{equation}\label{eq:quadratic}
		0 = g_f^2 + \( k_D + \frac{1}{|\Omega|} \int_\Omega \lc\cc +\ld\cd dx -  \Gamma \)g_f - \Gamma\,.
	\end{equation}
		
\subsection{Fibroblast cells and the matrix remodeling subsystem}		
	We describe the remodelling of the ECM with two subsystems. In the first we introduce the transition/transdifferentiation of the CSCs to fibroblast cells, and in the second we model the reconstruction of the ECM by the fibroblast cells. Contrary to the more usual assumptions of \emph{spontaneous} or \emph{self-remodeling} of the ECM, we promote in this work the more biologically relevant fibroblast-driven matrix remodelling.

\paragraph{The transdifferentiation subsystem.}
	An potential initial distribution of fibroblast cells exists in the environment and is locally enriched by a transdifferentiation programme of CSCs to \emph{cancer associated fibroblast} cells \cite{Katsumo.2013}. In this work we do not make a distinction between the two types of fibroblast cells, or the two types of induced ECM, although they present different biochemical and mechanical properties. We leave this question for a future investigation. 
	
	We assume that, similarly to the wound healing case, the fibroblast cells are directed towards regions of the ECM that need reconstruction. We model this property with a haptotactic movement towards lower densities of the ECM, \cite{Kim.2009}. This assumption is in a strong contrast to the haptotactic motion of the DCCs and CSCs, which is oriented towards the higher ECM densities.
	
	For their proliferation we employ the positive part logistic function to account only for the competition for free space and neglect the overcrowding induced cell death. Since the fibroblast cells exhibit a much shorter life span than the ``immortal'' cancer cells, we include an \emph{apoptosis} term with constant rate. This is a simplified assumption but suffices for the needs of this work. Accordingly, the transdifferentiation subsystem reads as:
	\begin{equation}\label{eq:sub.trans}
	\left\{
	\begin{aligned}
	\frac{\pd \cc}{\pd t}
		&= \gray{diffusion} -\chi_S\nabla \cdot \(\cc \nabla v\) - \mtra \, c^\text{CSC}  + \gray{proliferation}\\
	\frac{\pd \cf}{\pd t}
		&= \gray{diffusion} +\chi_F\nabla \cdot \(\cf \nabla v\) + \mtra \,  c^\text{CSC}  + \mu_F \, \cf (1-\cd-\cc-\cf-v)^+ - \beta_F \, \cf
	\end{aligned}\,,\right.
	\end{equation}
	where $\cf$ represents the density of the fibroblast cells, $\chi_F$ their haptotactic sensitivity, $\mtra$ their transdifferentiation production coefficient, and $\mu_F$ and $\beta_F$ their proliferation and apoptosis/decay rate.

\paragraph{The matrix remodelling subsystem.}
	Contrary to the more usual assumption of spontaneous or self-remodelling of the ECM, we promote here the (more biologically relevant) \emph{fibroblast-driven} ECM-remodelling limited by the availability of the free space and encoded by a logistic term:
	\begin{equation}\label{eq:sub.remod}
		\left\{
		\begin{aligned}
			\frac{\pd \cf}{\pd t}
				&= \gray{diffusion} + \gray{haptotaxis} + \gray{transdiff.} + \gray{proliferation}\\
			\frac{\pd v}{\pd t}
				&=-\gray{proteolysis} + \mu_F\,\cf (1-\cd-\cc-\cf-v)^+\\
		\end{aligned}\,,\right.
	\end{equation}
	where we again consider the positive-part \emph{logistic volume filling} reaction term to account only for the remodelling of the matrix and not for its degradation. In fact,  we assume that the matrix is degraded only by the proteolytic proteins MMP. The effect of the fibroblast-driven matrix remodelling in comparison to the more usual self-remodelling can be seen in the Figure \ref{fig:justify.FIB}.
	
\subsection{The full cancer invasion system}\label{sec:invasion}
		
	In total, the complete ECM invasion model results from the concatenation of the subsystems \eqref{eq:sub_basic}, \eqref{eq:sub_emt}, \eqref{eq:sub.trans}, and \eqref{eq:sub.remod} and reads as follows:
	\begin{equation}\label{eq:full.system}
		\begin{aligned}
			\frac{\pd \cd}{\pd t}
				&= D_D \Delta \cd -\chi_D\nabla \cdot \(\cd \nabla v\) & &- \memt \, \cd 
						& &+ \mu_D \, \cd\, (1-\cd-\cc-\cf-v)^+\,,\\
			\frac{\pd \cc}{\pd t}
				&= D_S \Delta \cc -\chi_S\nabla \cdot \(\cc \nabla v\) & - \mtra \,  \cc &+ \memt \, \cd
						& &+ \mu_S \, \cc\, (1-\cd-\cc-\cf-v)^+\,, \\
			\frac{\pd \cf}{\pd t}
				&= D_F \Delta \cf +\chi_F\nabla \cdot \(\cf \nabla v\) & + \mtra\, \cc & & -\beta_F \,\cf  
						& + \mu_F \, \cf\, (1-\cd-\cc-\cf-v)^+ \,, \\
			\frac{\pd v}{\pd t}
				&=  & & &  -\delta_v\, m v 
						& + \mu_v\, \cf\, (1-\cd-\cc-\cf-v)^+\,,\\
			\frac{\pd m}{\pd t}
				&=D_m\Delta m  &+ \alpha_D \,\cd &+ \alpha_S\,\cs & -\beta_m \,m\,.
		\end{aligned}
	\end{equation}
	This system is complemented by appropriate initial conditions and no-flux boundary conditions:
	\begin{equation}\label{eq:BCs}
		-D_D\frac{\pd \cd}{\pd \bf n} + \chi_D \cd\frac{\pd v}{\pd \bf n}=
		-D_S\frac{\pd \cc}{\pd \bf n} + \chi_S \cc\frac{\pd v}{\pd \bf n}=
			\frac{\pd m}{\pd \bf n}=0\,,
	\end{equation}
	where $\bf n$ is the outward unit normal vector to the computational domain $\Omega$. The model parameters that can be found in Table \ref{tbl:params}. They follow from investigations in the literature, e.g., \cite{KF.2010,Andasari.2011,chaplain5}, and from our  numerical experimentation, and reflect the current biological understanding of the problem. 		

\section{Mathematical methods}\label{sec:numerics}

In this section we present two mathematical techniques that we apply in order to investigate the dynamical behaviour of \eqref{eq:full.system} and \eqref{eq:BCs} and to predict the evolution of cancer cells. First we briefly describe the main components of our higher order implicit-explicit finite volume numerical scheme. In the second part we explain the way we compute parameter sensitivities.

\subsection{Numerical method}\label{sec:num.meth}
It is well-known that the solutions of the macroscopic cancer invasion systems typically develop clusters of cancer cell concentrations. The emerging of new clusters in time as well as their movement and interaction with each other makes the numerical treatment particularly challenging. The partial differential equation system \eqref{eq:full.system} that we consider can be classified as an advection-diffusion-reaction system. A wealth of numerical methods for these kind of systems in biological applications has been studied, for example, in \cite{hundsdorfer2003numerical,gerisch2006robust,Sfakianakis.2014b, Lukacova.2012}.

For the numerical simulations in this work we apply the finite volume method which, we have developed in \cite{Sfakianakis.2014b,Kolbe.2013}. We first discretise in space to derive a system of ordinary differential equations 
\begin{equation}\label{eq:num.ode}
	\frac{\partial \bf w_h}{\partial t} = \mathcal{A}(\bf w_h) + \mathcal{D}(\bf w_h) + \mathcal{R}(\bf w_h)\,,
\end{equation}
where $\bf w_h$ denotes a vector carrying the values of a piecewise constant function on $N$ spatially uniform computational cells for all components of the system.
The vector valued operators $\mathcal{A},~\mathcal{D},$ and $\mathcal{R}$ approximate the advection (i.e. haptotaxis), diffusion, and reaction terms of the system for every computational cell and every component of a finite volume solution $\bf w_h$.

We choose the diffusion and reaction operators in a conventional way. Central differences approximate the second derivative in the diffusion operator $\mathcal{D}$ and reaction terms are directly evaluated at the cell centers in the reaction operator $\mathcal{R}$. In order to compute fluxes for haptotaxis in a given cell we employ the dimensional splitting. Therefore we address briefly the advection operator $\mathcal{A}$ only in one-dimension. Let $c_i$ be the density of a migrating cell type $c$ (DCCs, CSCs, or fibroblasts) in the computational cell $C_i$ and $c_{i+1},~ c_{i-1}$ the densities in the adjacent computational cells. We consider the approximation,
 \begin{equation}
 \int_{C_i} \chi\, \nabla \cdot (c \nabla v) \approx a_{i+1/2}\, c_{i+1/2} - a_{i-1/2}\, c_{i+1/2}\,,
 \end{equation}
where $a_{i\pm 1/2}$ refers to the characteristic velocities, and $c_{i\pm1/2}$ to the reconstruction of cell densities at the interfaces of the computational cell.

The characteristic velocities in the haptotaxis case are proportional to the gradients of extracellular matrix density $v$. These are approximated by two point central differences. For the interface reconstructions we compute slopes, using the \textit{minimized-central} (MC) limiter \cite{van1977towards},
\begin{equation}
	s_i = \operatorname{minmod}\left( c_i - c_{i-1},\, \frac14(c_{i+1}-c_{i-1}),\, c_{i+1}-c_i\right)\,, 
\end{equation}
where
\begin{equation*}
	\quad \operatorname{minmod}(a, b, c) =  \begin{cases*} 
	\min \left\{ |a|, |b|, |c| \right\}~ \sign(b), & if $a,b,c > 0$, or $a, b, c < 0$\, ,\\
	0, & otherwise\, .
	\end{cases*}
\end{equation*}
Afterwards, we apply upwinding and define the reconstructed cell interfaces values
\begin{equation}
	c_{i+1/2} = c_{i+1-1/2} =
	\begin{cases}
		c_i + s_i, &\text{if }a_{i+1/2} > 0\,,\\
		c_{i+1} - s_{i+1},&\text{otherwise}\,.
	\end{cases}
\end{equation}

The resulting system of ODEs \eqref{eq:num.ode} is stiff due to the approximate diffusion operator. To reduce the number of time steps, which would be required by an explicit method, we employ implicit-explicit (IMEX) Runge Kutta schemes 
\cite{pareschi2005implicit, christopher2001additive}. The time integration scheme reads as follows
\begin{equation}
{\bf W_i} = {\bf w_h^n} + \Delta t^n \sum_{j=1}^{i-1}a_{ij}^E \, (\mathcal{A} + \mathcal{R})({\bf W_j}) + \Delta t^n \sum_{j=1}^{i} a_{ij}^I \, \mathcal{D}({\bf W_j}), \quad i=1,\dots,k+1\,, \quad {\bf w_h^{n+1}} = {\bf W_{k+1}}\,,
\end{equation}
thus we apply an explicit Runge Kutta method ($a_{ij}^E$) to the advection and reaction terms coupled with a diagonally implicit method ($a_{ij}^I$) for the diffusion terms. The Butcher tableau we employ has been derived in \cite{christopher2001additive}, and it yields a 4-stage scheme that is 3rd order accurate in time. The implicit treatment of diffusion allows us to use time steps $\Delta t^n$ proportional to the diameter of the computational cells, and the Courant-Friedrichs-Lewy stability number $0.49$. Note that in the case of the system \eqref{eq:full.system} our method is positivity preserving, i.e. given non-negative initial data, the numerical solution stays non-negative. We have already successfully applied this numerical method to study a chemotaxis-haptotaxis cancer system that promotes the role of \emph{urokinase} in the invasion of the ECM, and a preliminary version of the model \eqref{eq:full.system}, see  \cite{Sfakianakis.2014b, Sfakianakis.2015}. 

In Table \ref{tbl:conv} we demonstrate the second order \textit{experimental order of convergence} (EOC) of our numerical method in one- and two-space dimensions. To this end we have consecutively refined the grid globally and compared the numerical solutions. The experimental order has been computed using $EOC(N) = \log_2( E(N)-E(N/2))$, where $E(N)$ is the numerical error on a grid with $N$ mesh cells in each space dimension. The convergence of the method justifies that the phenomena we exhibit in Section \ref{sec:discussion} are no numerical artifacts but rather results of our mathematical model. 
				
\subsection{Parameter study}\label{sec:par.study}
	In phenomenological models like \eqref{eq:full.system} the dependence of the system dynamics on changes of the parameters is very interesting. Applying a local parameter analysis we can deduce the influence, that each part of the model has in the invasion process of the ECM by the cancer species. Accordingly, the parameters can be adjusted to better fit the model to realistic experimental data.

	For this process we recompute numerical solutions for the one-dimensional Experiment \ref{exp:1D.main}, see Section \ref{sec:discussion}, at the time instance $t=15$, varying one parameter at time and measure real valued functionals $\mathcal{F}$ of the solution with respect to the parameter changes. We are particularly interested in the mass of the DCCs and CSCs over the whole domain $\Omega$, the distance from the propagating CSC front to the slower DCC front, which quantifies the invasiveness of the CSCs, and the concentration of mesenchymal cells in the invading front.
	
	For a more comprehensive overview of the influence of all model parameters, we have computed the parameter sensitivity of the aforementioned functionals. This means that we compute approximate derivatives of the functionals with respect to the parameters. Again, we consider here Experiment \ref{exp:1D.main} at $t=15$. For feasibility we employ for each parameter $p$ central differences around the corresponding value $p_0$ from Table \ref{tbl:params} with a relative step size of $\delta p = p_0 \times 10^{-2}$ to compute the sensitivities
	\begin{equation}
		\mathcal{S}_p^\mathcal{F} = \frac{\mathcal{F}({\bf w_h}^{t=15}(p = p_0 + \delta p))- \mathcal{F}({\bf w_h}^{t=15}(p = p_0 - \delta p))}{2 \delta p}\approx \frac{\partial}{\partial p} \mathcal{F}({\bf w_h}^{t=15})\,,
	\end{equation}
	where $\mathcal{F}$ is the functional of interest, and ${\bf w_h}^{t=15}$ is the numerical solution (all components) of system \eqref{eq:full.system}, at time instance $t=15$. 
	
	In Table \ref{table:sensitivity} we present the results of this computation. Positive sensitivities  $\mathcal{S}_p^\mathcal{F}$ indicate an increase of the functional $\mathcal{F}$ (in the corresponding column) when increasing the parameter $p$ (in the corresponding line), while negative sensitivities indicate a decrease of the functional when increasing $p$. The absolute value quantifies the parameter influence on the corresponding attributes. Note that the sensitivity shows the effect, when increasing the parameter by one unit. Zero sensitivities in the column for the front distance mean that no influence on the relative invasion speed could be measured in the simulation. For a discussion of the results of the parameter study we refer to the following section.
	\begin{table}\small
	\renewcommand{\arraystretch}{1.3} 
	\begin{center}\begin{tabular}{r l|r|r|l}
		\multicolumn{2}{c|}{parameter}&  bio. relevant value & rescaled value & reference \\ \hline
	 	$D_D$ & diffusion coeff. of DCCs & $3.5 \times 10^{-10}\,\cm^2 \seconds^{-1}$ & $3.5 \times 10^{-4}$&\cite{bray2001cell, Anderson.2000, Chaplain.2005} \\ 
		$D_S$ & diffusion coeff. of CSCs & $3.8 \times 10^{-11}\,\cm^2 \seconds^{-1}$& $3.8 \times 10^{-5}$  & \cite{bray2001cell, Andasari.2011},  our choice \\ 
		$D_F$ & diffusion coeff. of fibroblasts & $3.5 \times 10^{-10}\,\cm^2\seconds^{-1}$ & $3.5 \times 10^{-4}$ &  \cite{KF.2010},  {our choice}\\ 
		$D_m$ & diffusion coeff. of MMPs & $2.5 \times 10^{-9}\,\cm^2 \seconds^{-1}$ & $2.5 \times 10^{-3}$ &\cite{robbins1965further, Stokes419, Chaplain.2005} \\  \hline
		$\chi_D$ & haptotaxis coeff. of DCCs & $1.3\,\cm^{d+2} \mol^{-1} \seconds^{-1}$ & $8\times 10^{-3}$ & \cite{Stokes419, Chaplain.2005}\\
		$\chi_S$ & haptotaxis coeff. of CSCs & $62.5\,\cm^{d+2} \mol^{-1} \seconds^{-1}$ &$4\times 10^{-1}$ & \cite{Stokes419, Andasari.2011}, our choice\\
		$\chi_f$ & haptotaxis coeff. of fibroblasts & $1.3\,\cm^{d+2} \mol^{-1} \seconds^{-1}$ & $8\times 10^{-3}$ & \cite{Stokes419}, our choice\\ \hline
		$\ld$ & EGF receptors per DCC & $1.9 \times 10^{7}$ &$1$ & {\cite{imai1982epidermal, ozcan2006nature, klein2004structure}, }our choice\\  
		$\lc$ & EGF receptors per CSC & $2.7 \times 10^{7}$ & $1.4$& {\cite{imai1982epidermal, ozcan2006nature, klein2004structure}, }our choice\\ \hline
		$k_D$ & EGF unbinding/binding & $3.2 \times 10^{-12}\, \cm^{-d} \mol $ & 2 &  {\cite{ozcan2006nature}, }estimated\\ 
		$\Gamma$ & average of total EGF& $8 \times 10^{-14}\,\cm^{-d} \mol $& $0.05$ &  estimated\\ \hline
		$\mu_0 $ & EMT factor& $5.5 \times 10^{-6}\,\seconds^{-1}$ & 0.055&  estimated \\
		$\mu_{1/2} $ & critical EGF density& $3.2 \times 10^{-12}\, \cm^{-d} \mol$ & $2$ &  estimated\\ \hline
		$\mu_D $ & proliferation rate of DCCs& $2 \times 10^{-5}\,\seconds^{-1}$ & $0.2$ & \cite{stokes1991analysis, Chaplain.2005}\\ 
		$\mu_S $ & proliferation rate of CSCs& $10^{-5}\,\seconds^{-1}$ & $0.1$ &  our choice\\ 
		$\mu_F $ & proliferation rate of fibroblasts& $10^{-5}\,\seconds^{-1}$ & $0.1$ & our choice \\ 
		$\mu_v $ & ECM remodelling rate & $1.9 \times 10^{8}\,\seconds^{-1}$ & $25$ & our choice\\\hline
		$\beta_F $ & apoptosis rate of fibroblasts& $3 \times 10^{-7}\,\seconds^{-1}$ & $3 \times 10^{-3}$ & \cite{KF.2010}\\
		$\beta_m $ & decay rate of MMPs& $10^{-4}\,\seconds^{-1}$ & $1$ & \cite{Anderson.2000}\\ \hline
		$\alpha_D $ & MMP production rate of DCCs & $2.3\,\seconds^{-1}$ & $0.1$ & \cite{Anderson.2000,Chaplain.2005}\\
		$\alpha_S $ & MMP production rate of CSCs & $22.9 \,\seconds^{-1}$ & $1$ &  our choice\\ \hline
		$\delta_v $ & ECM degradation rate & $5.3 \times 10^{9}\, \cm^d \mol^{-1} \seconds^{-1}$ & $1$ & \cite{Anderson.2000,Chaplain.2005}\\ 
		$\mtra $ & transdifferentiation rate & $10^{-6}\,\seconds^{-1}$ & $0.01$ &estimated \\ \hline
		$c\rfr $ & reference cell density & $8.3 \times 10^{-20}\, \cm^{-d} \mol$ & - & \cite{KF.2010} \\
		$v \rfr $ & reference ECM density & $6.4 \times 10^{-9}\,\cm^{-d} \mol$& - & \cite{KF.2010}\\
		$m \rfr $ & reference MMP density & $1.9 \times 10^{-14}\,\cm^{-d} \mol$ & - & \cite{KF.2010}\\
		$g \rfr $ & reference EGF density & $1.6 \times 10^{-12}\,\cm^{-d} \mol$ & - & {\cite{jeulin2008egf, KF.2010}} \\ \hline
		$D$ & diffusion scaling coeff. & $10^{-6}\, \cm^2 \seconds^{-1}$ & - & \cite{bray2001cell, Chaplain.2005} \\
		$t_{sc}$ & time scaling coeff. & $10^{4}\, \seconds$ & - & \cite{Anderson.2000, Chaplain.2005} \\ \hline
	\end{tabular}\end{center}
	\caption{Parameter values in their derived biological units and in a rescaled formulation which we have used in our simulations. For the non-dimensionalization, the reference densities and the scaling coefficients in the lower part of the table have been used. The integer $d\in\{1,2\}$ denotes the space dimension. Parameter values referenced by ``our choice'' have been chosen according to our biological understanding of the processes, while we have decided on ``estimated'' parameter values after numerical experimentation.}\label{tbl:params}
	\end{table}

\begin{table}
	\resizebox{\linewidth}{!}{
	\begin{tabular}{ r| r r | r r}
		Grid	& $L^1$ error & EOC & $L^2$ error& EOC \\ \hline
			512/1024 & 1.953e-05  &   &3.199e-05  & \\
		1024/2048 & 4.430e-06  & 2.140   & 7.359e-06  & 2.120    \\
		2048/4096 & 1.061e-06  & 2.062   & 1.656e-06  & 2.152  \\
		4096/8192  & 2.569e-07  & 2.046   &3.857e-07  & 2.102   \\
	\end{tabular}\hspace{2em}
	\begin{tabular}{ r| r r   | r r }
		Grid	&  $L^1$ error & EOC &  $L^2$ error & EOC  \\ \hline
		16$\times$16/32$\times$32    & 1.020e-01    &       &3.025e-02      &     \\	  
		32$\times$32/64$\times$64  & 3.090e-02  &1.723   &1.023e-02  &1.564     \\
		64$\times$64/128$\times$128  & 8.453e-03  &1.870   &2.715e-03  &1.914   \\
		128$\times$128/256$\times$256 & 2.180e-03  &1.955   &6.945e-04  &1.967    \\	
	\end{tabular}}
	\caption{Experimental convergence rates of the numerical method presented in Section \ref{sec:num.meth} employed in the system \eqref{eq:full.system} for the DCCs in one-dimensional (left) and two-dimensional (right).  In the one-dimensional case, Experiment \ref{exp:1D.main} is performed for the final time $t=15$, and in the two-dimensional case, Experiment \ref{exp:2D.conv} is realized for the final time $t=10$. In both cases the results confirm the second order convergence of the numerical method.}\label{tbl:conv}
\end{table}	

\section{Experiments and discussion}\label{sec:discussion}

The initial conditions that we use in the one-dimensional case are:
\begin{equation}\label{eq:ic.1d}
\cd_0(x) = \exp (-20\,x^2)\,,\quad \cc_0(x) = \cf_0(x) = 0\,, \quad v_0(x) = 1- 0.5\,\cd_0(x)\,, \quad m_0(x) = 0.2\, \cd_0(x)\, ,
\end{equation}
for all $x\in \Omega$, whereas in the two-dimensional case:
\begin{equation}\label{eq:ic.2d.dcc}
	\cd_0(\mathbf x) = 
	\begin{cases}
		\sin\(5\,\arctan\(\frac{x_2}{x_1}\)\),&\text{if }x_1^2+x_2^2< 4,~ x_2\leq 0,~ x_1\geq 0\,,\\
		\sin\(5\,\(\pi + \arctan\(\frac{x_2}{x_1}\)\)\),&\text{if }x_1^2+x_2^2 < 4,~ x_2\leq 0,~ x_1< 0\,,\\
		0, & \text{otherwise}\,,
	\end{cases}
\end{equation}
\begin{equation} \label{eq:ic.2d.rest}
	\cc_0(\mathbf x) = \cf_0(\mathbf x)=0, \ v_0(\mathbf x) = 1 - \cd_0(\mathbf x), \ m_0(\mathbf x) = \frac{1}{20} \, \cd_0(\mathbf x), \quad \mathbf x=(x_1,x_2)\in \Omega\,.
\end{equation}	

	We use a fine computational grid with $7000$ cells in the one-dimensional cases and $250\times 250$ for the two-dimensional simulations. Further details on the various experimental settings are given in the Appendix.

	Figure \ref{fig:invasion1D} presents the numerical solution of the system \eqref{eq:full.system} over a one-dimensional domain with ICs \eqref{eq:ic.1d} and parameters taken from the Table \ref{tbl:params}. In the initial conditions no CSCs are included, rather they are formed by the DCCs after undergoing EMT. The CSCs exhibit higher motility and invasiveness than the DCCs, they escape the main body of the tumor and invade the ECM while developing highly dynamic phenomena like the merging and emerging of concentrations. In a similar way, Figure \ref{fig:invasion2D} presents the dynamics of the same model \eqref{eq:full.system} and parameters on a two-dimensional domain. We see in particular, laterally propagating waves in the migration of the CSCs, which are due to the non-convexity of the DCC front in the initial conditions. 

	\begin{figure}[t]
		\centering
		\resizebox{\linewidth}{!}{
		\begin{tabular}{c c c c}
			\subfigure[$t=1.1$]{\includegraphics{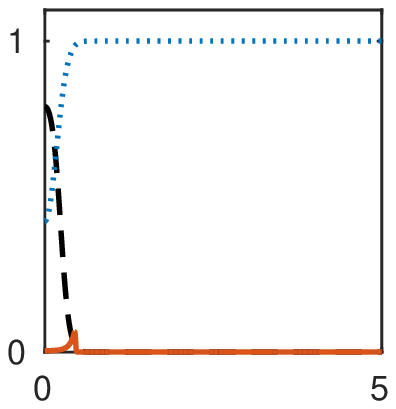}}
			&\subfigure[$t=5.1$]{\includegraphics{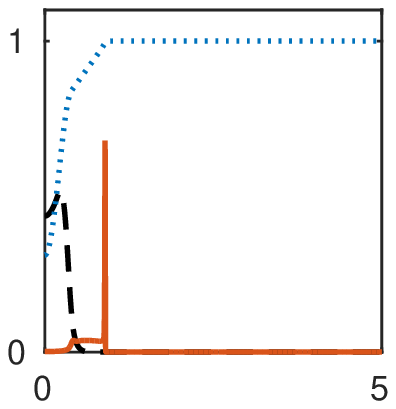}}
			&\subfigure[$t=12$]{\includegraphics{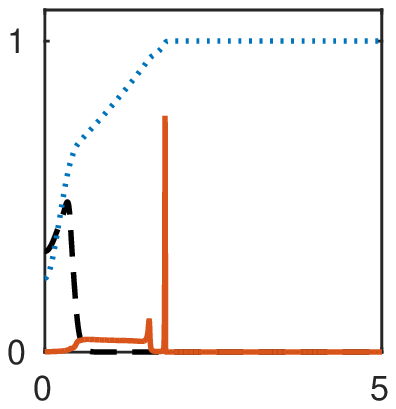}}
			&\subfigure[$t=18.1$]{\includegraphics{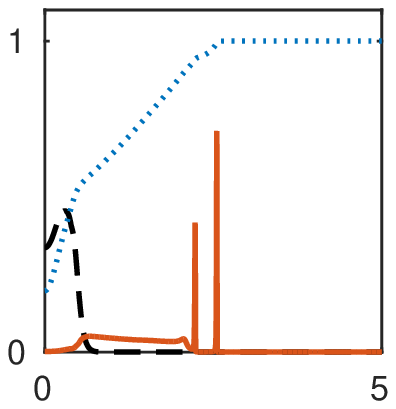}}\\
		
			\subfigure[$t=20.1$]{\includegraphics{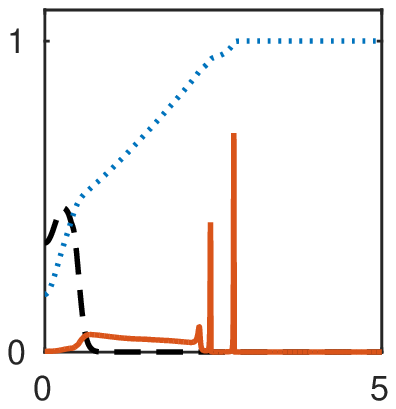}}
			&\subfigure[$t=24.5$]{\includegraphics{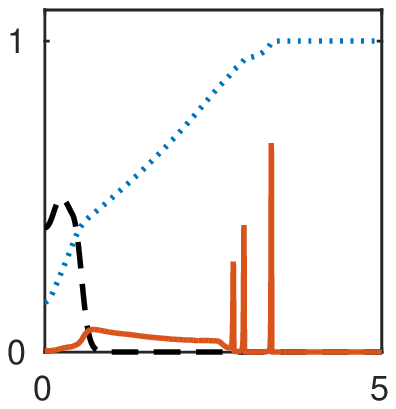}}
			&\subfigure[$t=27.9$]{\includegraphics{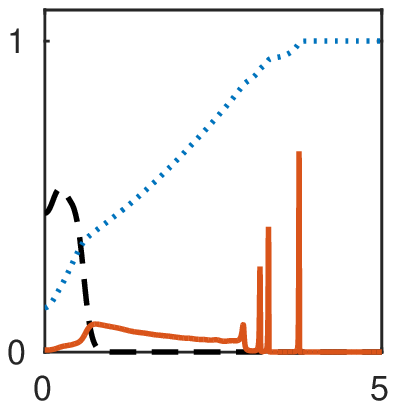}}
			&\subfigure[$t=32.1$]{\includegraphics{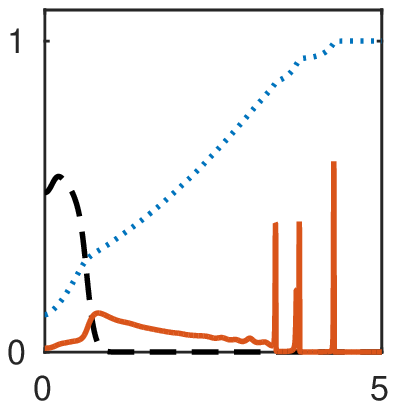}}
		\end{tabular}}
		\vfill \vspace{.25em}
		\includegraphics[scale=0.8]{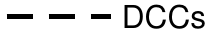}\quad\includegraphics[scale=0.8]{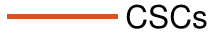}\quad
			\includegraphics[scale=0.8]{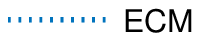}
				
		\caption{Time evolution of the one-dimensional cancer invasion system \eqref{eq:full.system}. (a) the CSCs are produced from the DCCs via the EMT,  (b)-(c) due to their higher motility, the CSCs escape the main body of the tumor and invade the ECM faster than the DCCs. Two distinct propagating fronts have been formed, (d)-(f) the CSCs present dynamic travelling merging and emerging concentrations; the DCCs propagate into the ECM slower. The computational setting is described in Experiment \ref{exp:1D.main} in the Appendix.}\label{fig:invasion1D}
	\end{figure}

	\begin{figure}[t]
		\resizebox{\linewidth}{!}{
		\begin{tabular}{c| c|c|c}
			&	DCC & CSC & ECM\\
			\rotatebox{90}{\hspace{1.5cm}$t = 30$}
			&\includegraphics{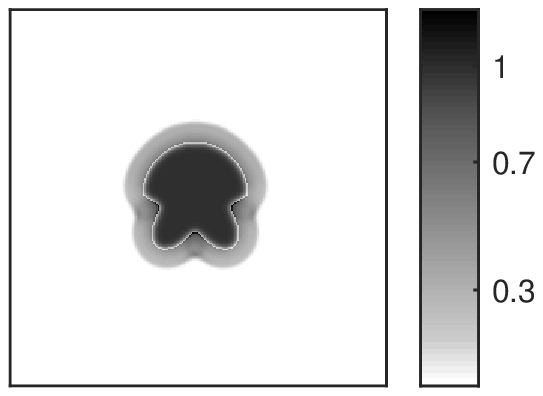} 
			&\includegraphics{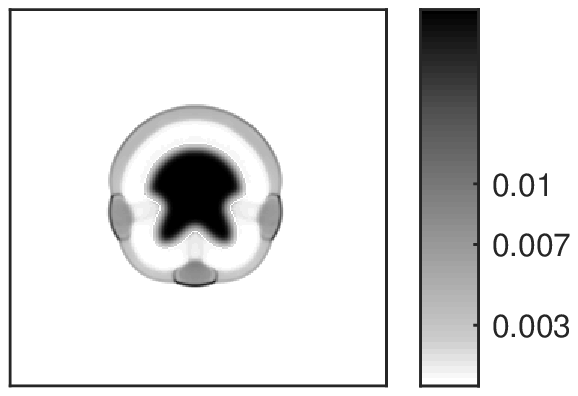}
			&\includegraphics{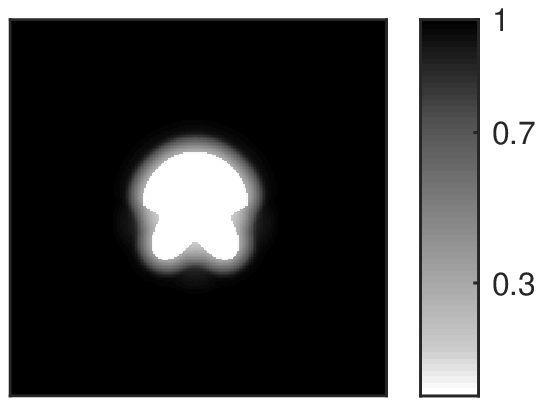} \\
			\hline
			\rotatebox{90}{\hspace{1.5cm}$t = 55$}
			&\includegraphics{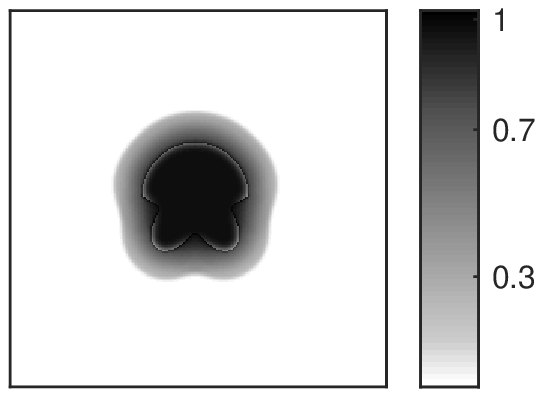} 
			&\includegraphics{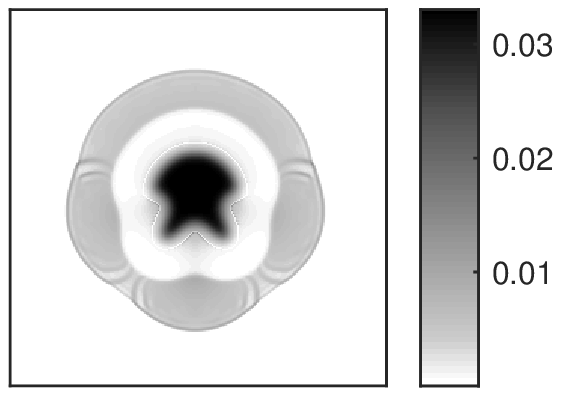}
			&\includegraphics{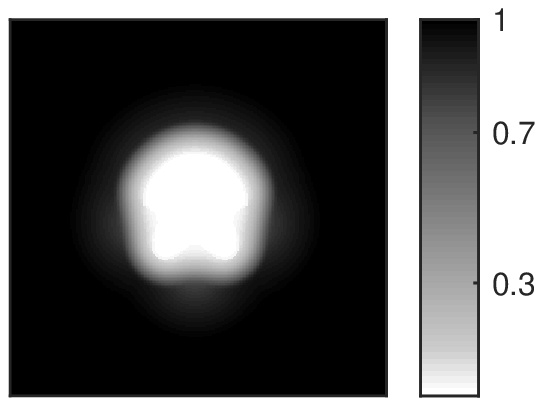} \\
			\hline
			\rotatebox{90}{\hspace{1.5cm}$t = 80$}
			&\includegraphics{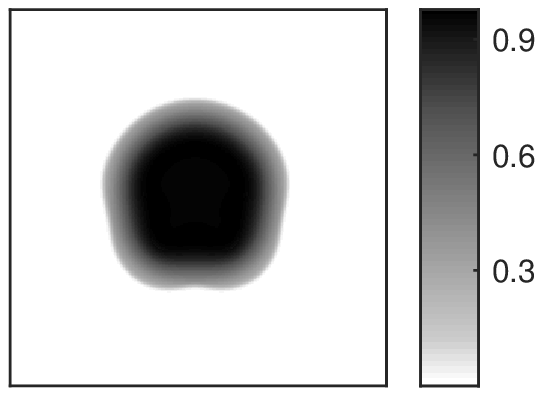} 
			&\includegraphics{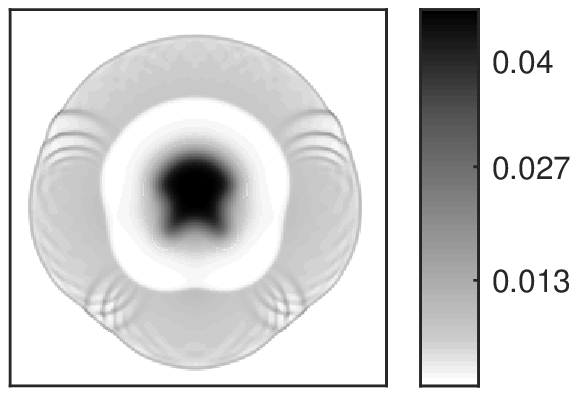}
			&\includegraphics{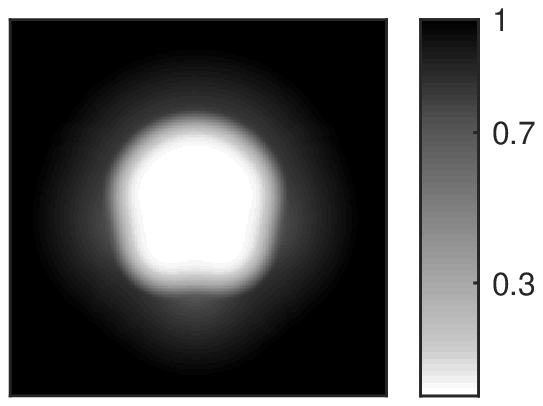}
		\end{tabular}}
		\caption{Time evolution of the two-dimensional cancer invasion system \eqref{eq:full.system}, in Experiment \ref{exp:2D.main}. Similar to the one-dimensional case (Figure \ref{fig:invasion1D}), the invasion of the DCCs onto the ECM is smooth and mostly driven by proliferation and diffusion. The CSCs exhibit richer dynamics and patterns in their invasion. In particular, laterally propagating and interacting waves appear due to the non-convexity of the front of the initial concentrations. The ECM is depleted by the MMPs which in turn are produced by the cancer cells and is constantly remodelled by the fibroblast cells. An initially uniform distribution of the ECM is assumed. 
		}\label{fig:invasion2D}	
	\end{figure}
	
\begin{figure}[t]
		\resizebox{\linewidth}{!}{
		\begin{tabular}{c| c|c|c}
			&	DCC & CSC & ECM\\
			\rotatebox{90}{\hspace{.8cm}no remodelling}
			&\includegraphics{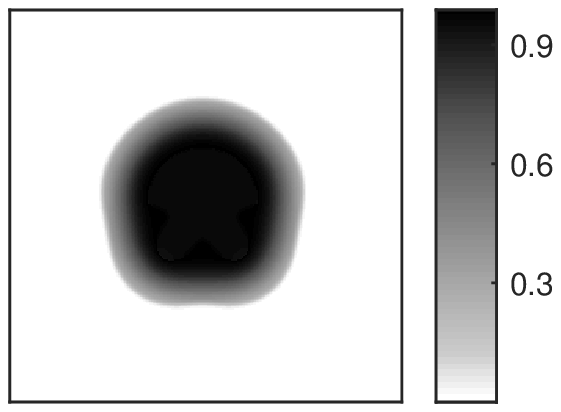} 
			&\includegraphics{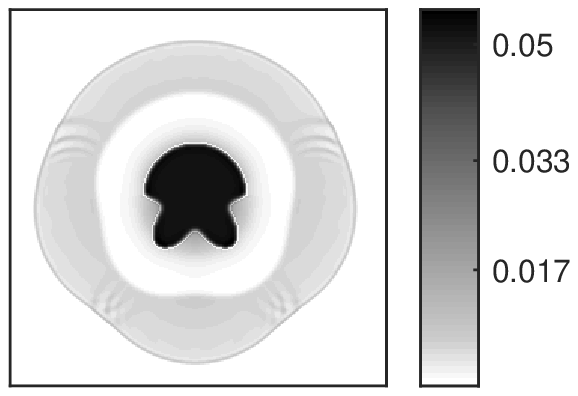}
			&\includegraphics{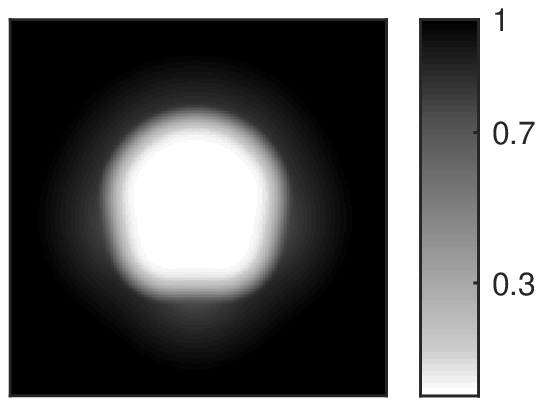} \\
			\hline
			\rotatebox{90}{\hspace{.8cm}self remodelling}
			&\includegraphics{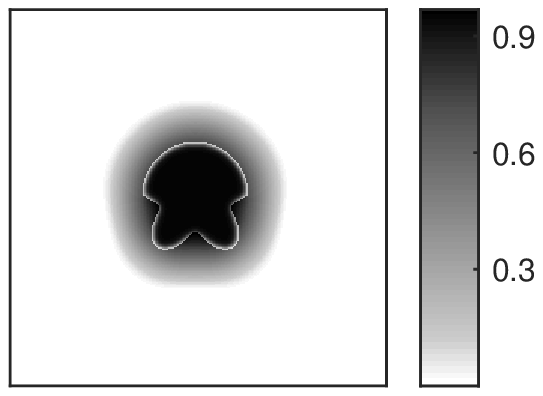} 
			&\includegraphics{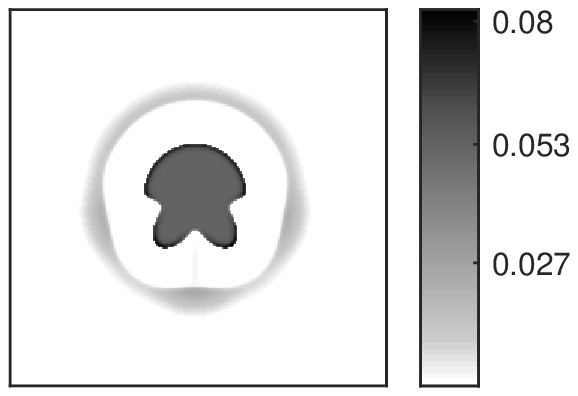}
			&\includegraphics{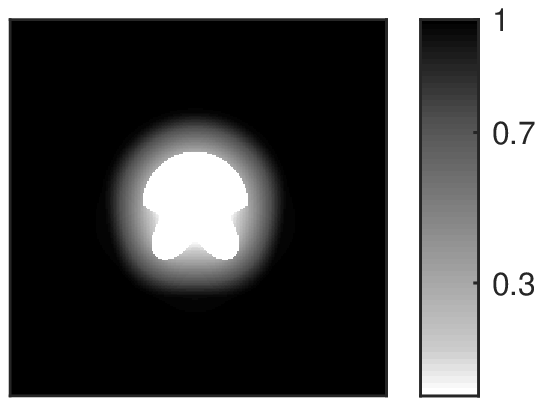} \\
			\hline
			\rotatebox{90}{\hspace{.4cm}fibroblast remodelling}
			&\includegraphics{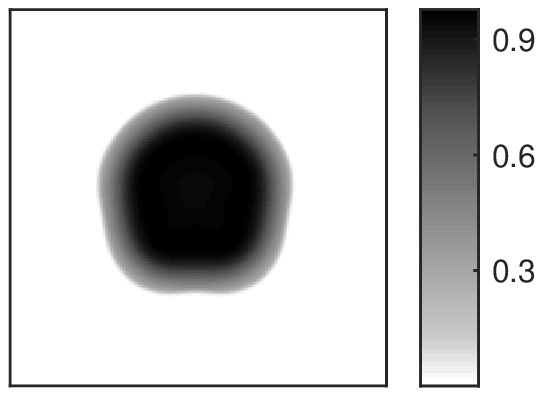} 
			&\includegraphics{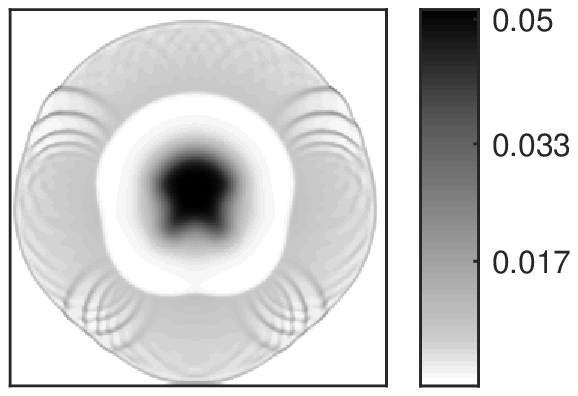}
			&\includegraphics{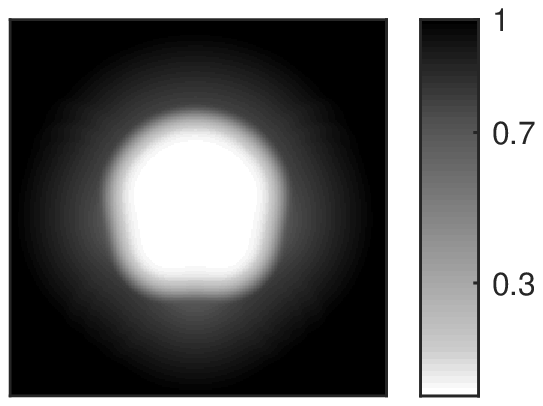}
		\end{tabular}}
		\caption{Comparison between three different types of ECM remodelling at time $t=90$. (Upper row, Experiment \ref{exp:2D.no.remod}) The ECM is depleted by the MMPs and is not remodelled. Newly sprout CSCs in the kernel of the tumour are unable to adhere and escape the main body of the tumour. This case is used for comparison reasons. (Middle row, Experiment \ref{exp:2D.self.remod}) The ECM is automatically recovered as long as there is some pre-existing ECM. A non-uniform ECM is remodelled by filling the free volume leading to a uniform ECM distribution. (Lower row, Experiment \ref{exp:2D.main}) The matrix is remodelled by fibroblast cells that are haptotactically directed towards lower densities of the ECM. This recovers the matrix in the kernel of the tumour allowing for the CSCs to adhere and escape the tumour. 
		}\label{fig:justify.FIB}	
	\end{figure}

	In Figure \ref{fig:justify.FIB} we compare three different types of ECM remodelling: the case without matrix remodelling (Experiment IIa) is used for comparison reasons. The matrix is depleted by the MMPs, which is produced by the cancer cells, creating ``holes''in which the cells cannot attach and translocate. The case of self remodelling is included as it is one of the most prominent in the literature. We see the stronger confinement of both types of cancer cells and the ``sharper'' ECM. In the cancer-associated fibroblast remodelling case we notice the faster invasion of the CSCs, as well as repair of the ECM by the fibroblast cells.
		
\begin{figure}[t]
		\centering
		\resizebox{\linewidth}{!}{
			\begin{tabular}{c|c c c c l}
				 &$t = 4$ & $t=14$ & $t=24$ & $t=32$\\ \hline
					\rotatebox{90}{\hspace{1.5em}constant rate EMT}
					&\includegraphics{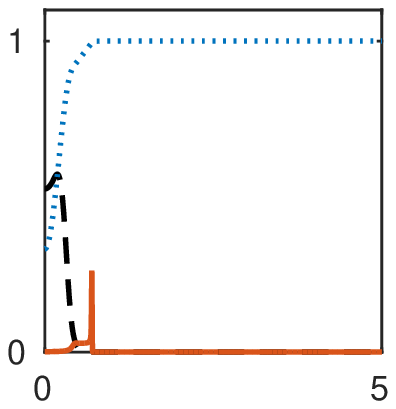}
					&\includegraphics{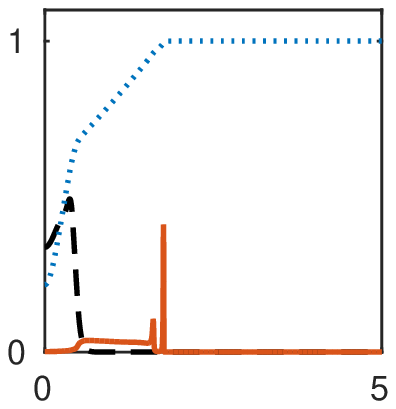}
					&\includegraphics{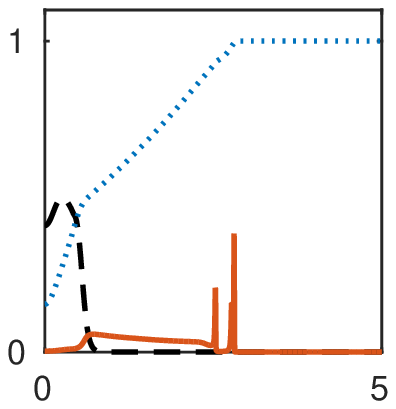}
					&\includegraphics{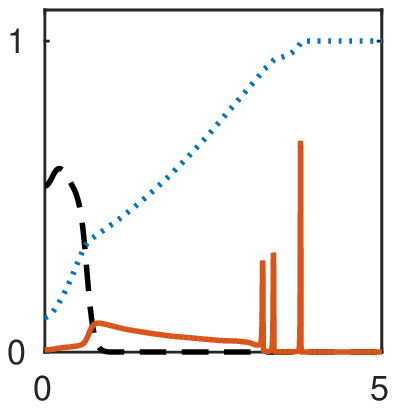}
					\\ \hline
					\rotatebox{90}{\hspace{1.8em}EGF-driven EMT}
					&\includegraphics{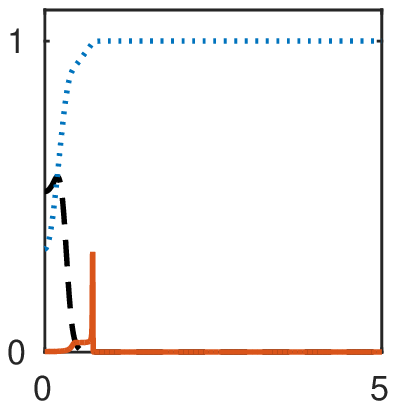}
					&\includegraphics{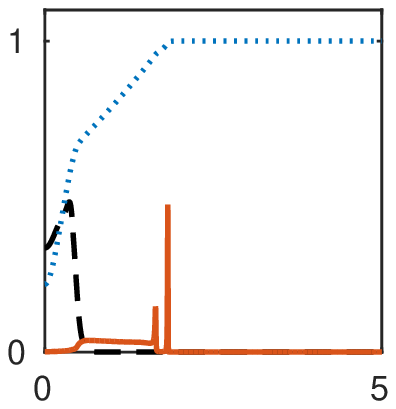}
					&\includegraphics{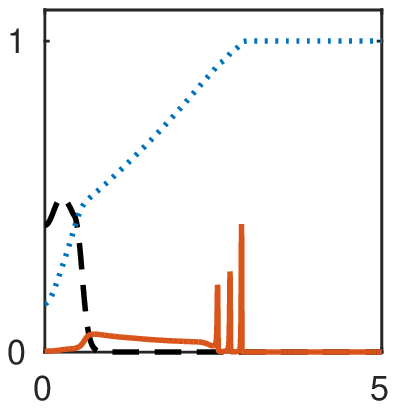}
					&\includegraphics{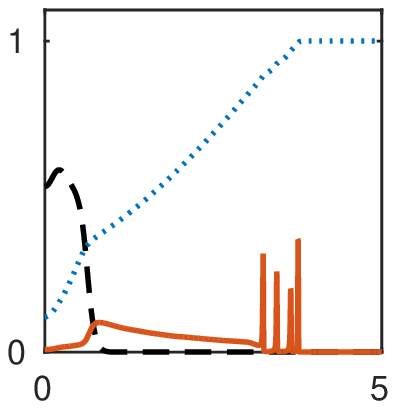}
					\\ \hline
					\rotatebox{90}{\hspace{2.5em}reduced EGF}
					&\includegraphics{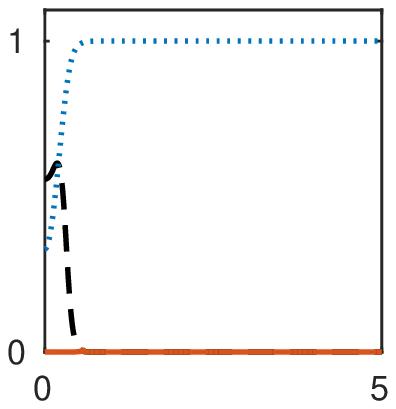}
					&\includegraphics{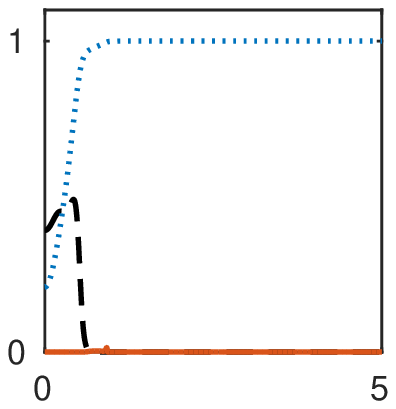}
					&\includegraphics{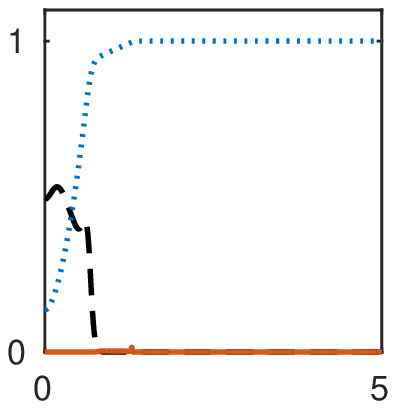}
					&\includegraphics{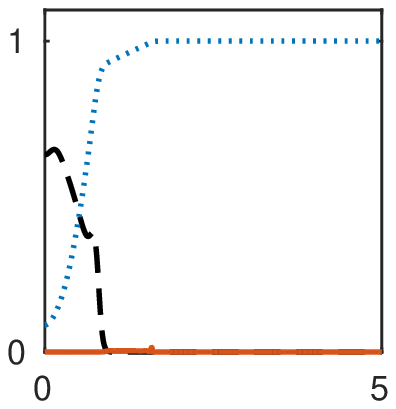}
			\end{tabular}}
			\vfill \vspace{.25em}
		\includegraphics[scale=0.8]{legend-dcc-exp-1d}\quad\includegraphics[scale=0.8]{legend-csc-exp-1d}\quad
		\includegraphics[scale=0.8]{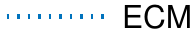}
		\caption{Comparison between the constant rate EMT (upper row, Experiment \ref{exp:1D.ctEMT}), versus the EGF-driven EMT, i.e. $\memt$ given by \eqref{eq:muemt_2} (middle and lower rows,  Experiments \ref{exp:1D.var.EMT}, \ref{exp:1D.var.EMT.low.EGF}). The parameters in the upper and middle graph are based on the Table \ref{tbl:params} and are adjusted so that the invasiveness of the CSCs and the DCCs coincide. We observe that the components of the solution are qualitatively the same; there is a difference in the number of peaks that the CSCs exhibit. In the third row, we present a particular experiment of the EGF-driven EMT case, where the total EGF has been depleted to a low value ($10^{-3}$). A single propagating low-density peak of CSCs appears, demonstrating the adjustability of our model to the  pathological situation of very low EGF density.
		}\label{fig:compEMT}
	\end{figure}

	In Figure \ref{fig:compEMT} we compare different EMT strategies: we compare a typical constant EMT rate $\memt$ in \eqref{eq:full.system} and our EGF-driven $\memt$ coefficient proposed in \eqref{eq:muemt_2}. For the one-dimensional case we start this comparison with parameters from Table \ref{tbl:params} (modified to fit the one-dimensional case), and subsequently adjusted it so that the invasiveness (propagating front of the CSCs and DCCs) and distributions of the cancer cells coincide at different time instances for the constant rate and the EGF-driven EMT, see Figure \ref{fig:compEMT} upper and middle rows. In the lower row we consider the EGF-driven EMT and the pathological situation, where the total amount of EGF given by $\Gamma$ in \eqref{eq:egf_mass} is extremely low (of the order of $10^{-3}$). As expected only a small amount of DCCs have undergone EMT, leading to a small but still existent number of CSCs which invade the ECM.
	
\begin{figure}[t]
		\centering
		\resizebox{\linewidth}{!}{
		\begin{tabular}{c c}
				EGF-driven EMT & constant rate EMT\\
			\includegraphics{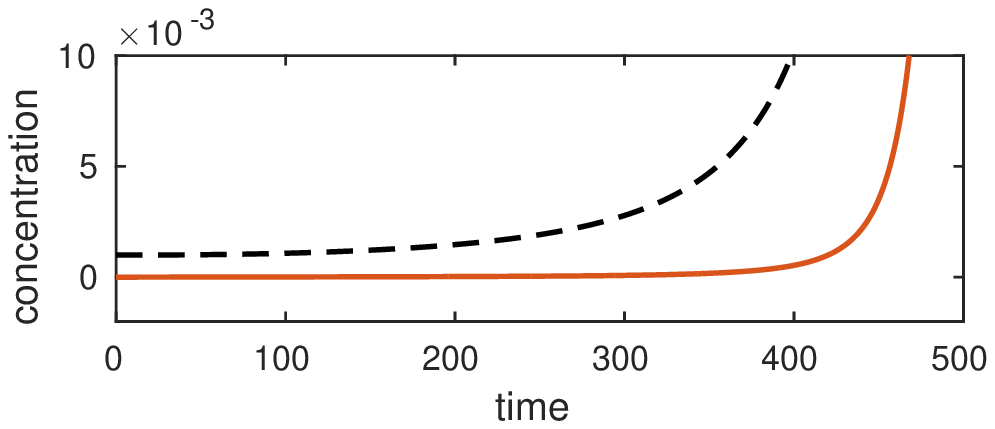}	&\includegraphics{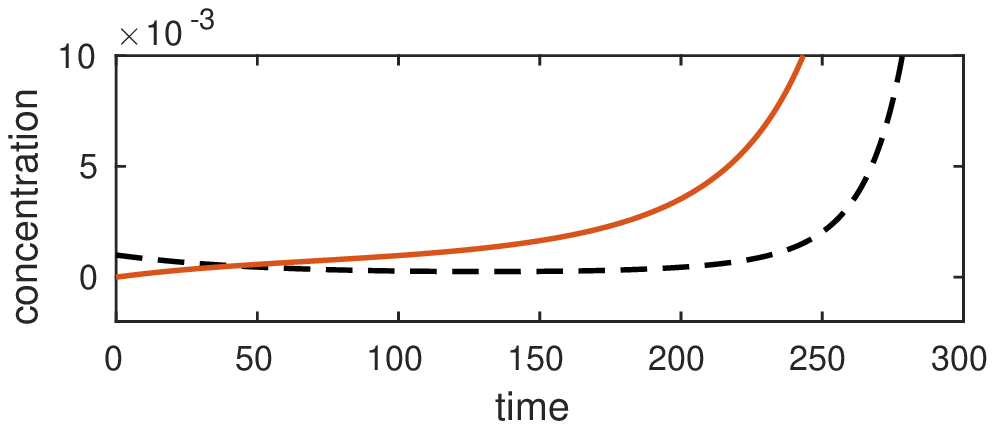}
		\end{tabular}}
		\vfill \vspace{.25em}
		\includegraphics[scale=0.8]{legend-dcc-exp-1d}\quad \includegraphics[scale=0.8]{legend-csc-exp-1d}
		
			\caption{A comparison of the EGF-driven EMT (left, Experiment \ref{exp:1D.var.EMT.small.IC}) with the constant rate EMT (right, Experiment \ref{exp:1D.ct.EMT.small.IC}) in terms of the production of CSCs. We use the same parameters as in Figure \ref{fig:compEMT} with an initial DCC concentration of a maximum value $10^{-3}$. We can see that in the EGF-driven EMT case there is no concentration of the CSCs comparable to the one of the DCCs until the DCC concentration has increased considerably. Loosely translated, our model predicts that small tumours produce CSCs that cannot be detected due to their low densities. On the other hand, in the constant rate EMT case the density of the CSCs raises quickly to become comparable to the DCCs level and can be easily detected.
			}\label{fig:compEMTODE}
	\end{figure}

	In Figure \ref{fig:compEMTODE} we go one step further and first consider the same coefficients as the ones that give the coincidence between the constant rate and the EGF-driven EMT, see Figure \ref{fig:compEMT}. We consider a small initial tumour (with a maximum density of $10^{-3}$) and simulate the evolution of the cancer dynamics. We see that in the constant rate EMT case, the CSCs quickly rise and achieve densities comparable to the DCCs'. This means that the metastatic part of the tumour would be as detectable as its main body. In contrast, in the EGF-driven EMT case, we observe that the density of the CSCs remains low (negligible relatively to the DCCs for a long time), but start to increase later. But even in the initial phase a few CSCS are present (relative number at t= 400 or so), and thus can invade the ECM. Our model hence is able to produce a scenario where metastatic CSCs can be produced also by ``small'' tumours and remain un-detectable in comparison to the main body of the DCCs.
	
	\begin{figure}[t]
		\resizebox{\linewidth}{!}{
		\begin{tabular}{c c}
			\includegraphics{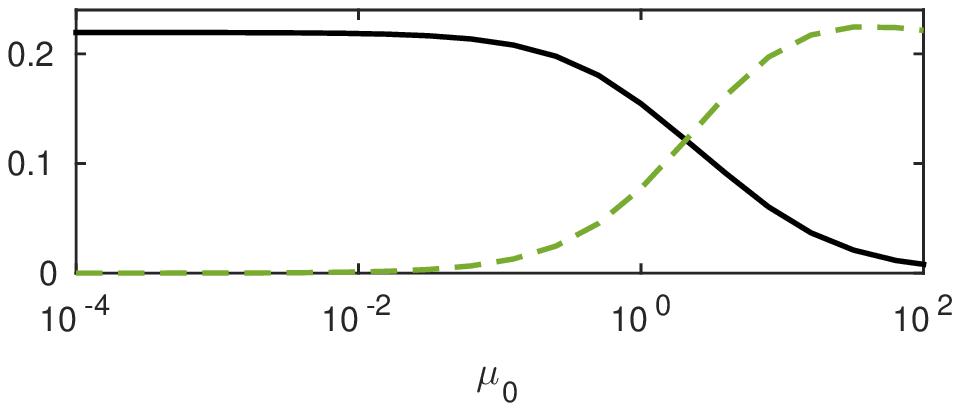} & \includegraphics{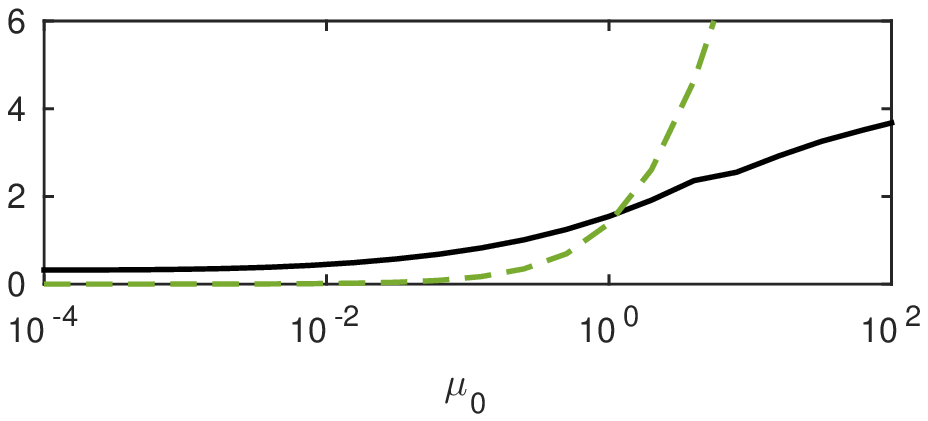} \\ 
			\includegraphics{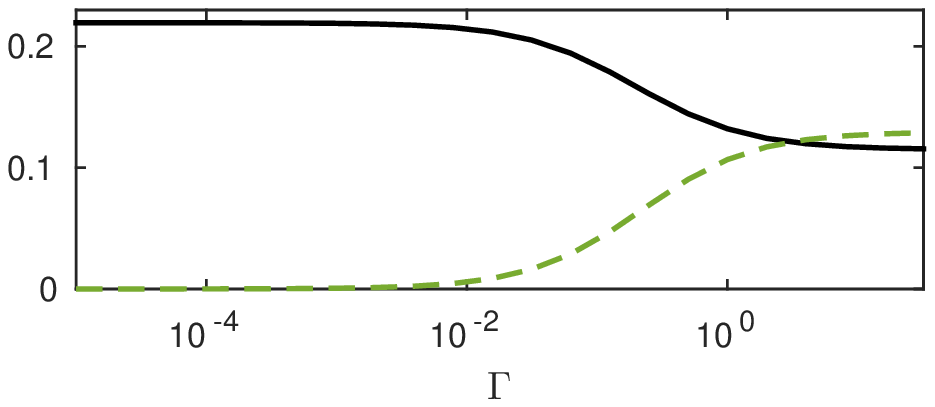} & \includegraphics{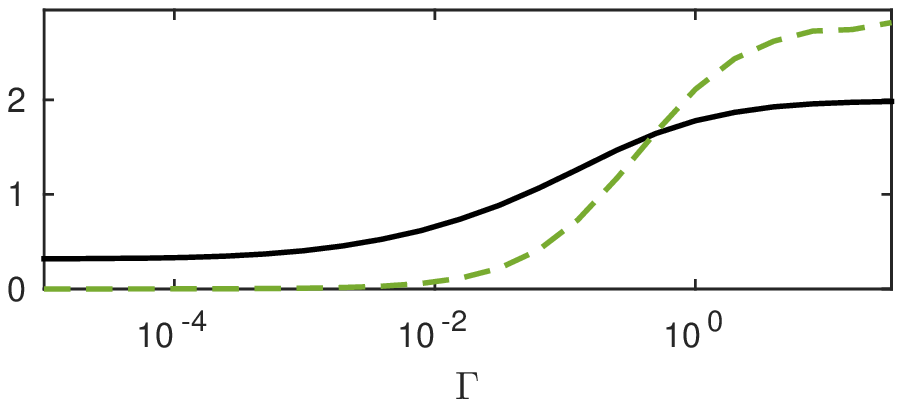} \\ 
			\includegraphics{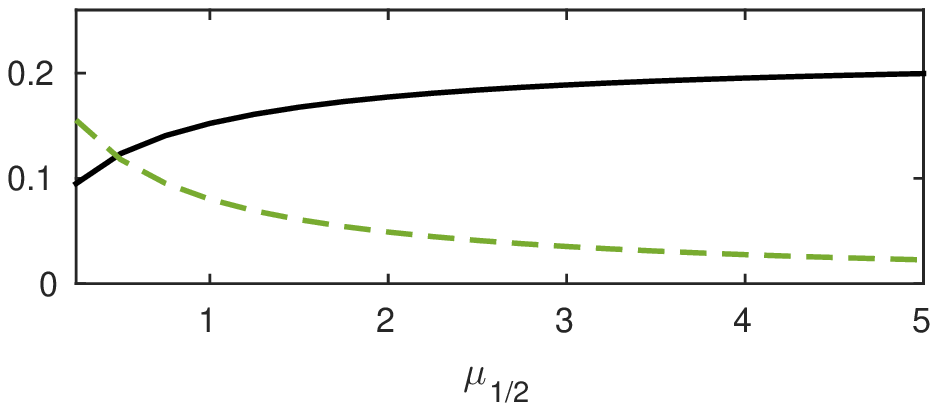} & \includegraphics{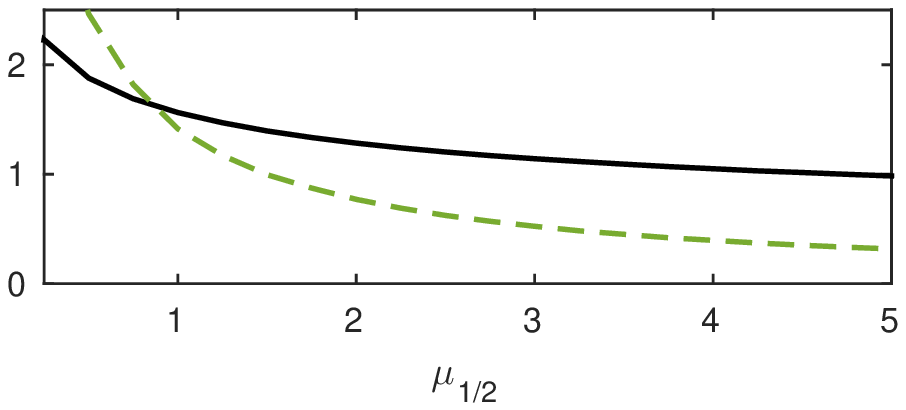} \\ \\
			\includegraphics{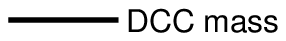}\quad
			\includegraphics{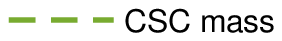} 
			&\includegraphics{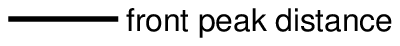}\quad
			\includegraphics{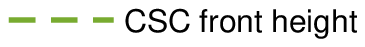}
		\end{tabular}}
			\caption{The influence of the parameters $\mu_0$, $\Gamma$, $\mu_{1/2}$ on the mass and the invasiveness of the cancer cells. (Upper row) High values of the maximum EMT rate $\mu_0$ in \eqref{eq:muemt} promote the CSCs and deplete the DCCs (left). The front distance between the two families of cancer cells and the density of the CSCs at the propagating front increases (right). (Middle row) When the (average) mass of EGF $\Gamma$, cf. \eqref{eq:egf_mass}, increases, the masses of the two cancer cell types are driven to a stable situation (left). Similarly, their relative invasiveness and the density of the CSCs indicates stagnation (right). (Lower row) The increase of $\mu_{1/2}$ in \eqref{eq:muemt} is beneficial for the DCCs. It depletes the CSCs and decreases the density and the distance of the propagating fronts; in effect it weakens the migration and growth of CSCs. This graph suggests that  enhancing $\mu_{1/2}$ prevents the migration of the CSCs, cf. Experiment \ref{exp:1D.main}.} \label{fig:parameter.dependence}
	\end{figure}
	
\begin{table}\small
	\renewcommand{\arraystretch}{1.} 
	\begin{center}\begin{tabular}{ r l | c c c c } 
			\multicolumn{2}{c|}{parameter}&  DCC mass & CSC mass & front distance  & front height\\ \hline
			$D_D$ & diffusion coeff. of DCCs  &  +1.2342e+01   &   -8.1722e+00  &    +1.3714e+02  &    -7.3024e+01 \\
			$D_S$ & diffusion coeff. of CSCs  &   +9.4182e-02  &    -3.2915e-01 &     $\pm$0 &     -9.5376e+03 \\
			$D_F$ & diffusion coeff. of fibroblasts &   -2.2642e-02  &    +2.1300e-03 &     $\pm$0 &     +3.9125e-01 \\
			$D_m$ & diffusion coeff. of MMPs &   -2.9334e-01  &    -9.5195e-02 &     -2.8800e+01 &     -1.6365e+02 \\ &&& \\
			
			$\chi_D$ & haptotaxis coeff. of DCCs &  +8.9096e-01   &   -3.4507e-01   &   -1.3500e+01   &   -3.1985e+00 \\
			$\chi_S$ & haptotaxis coeff. of CSCs &   -9.4425e-03 & -3.5106e-03   &   +1.5600e+00   &   +4.9956e-01 \\
			$\chi_F$ & haptotaxis coeff. of fibroblasts &  -7.0482e-03  &    -8.6437e-04  &    $\pm$0  &    -1.6845e-03 \\	&&& \\
			
			$\ld$ & EGF receptors per DCC &  -3.2691e-02   &   +4.6710e-02  &    +5.4000e-01  &    +7.5617e-01 \\
			$\lc$ & EGF receptors per CSC &  +5.1239e-04   &   -6.5509e-04  &    $\pm$0  &    -1.4768e-02 \\ &&& \\
			
			$k_D$ & EGF unbinding/binding &		+5.1703e-02    &  -7.4019e-02    &  -8.6400e-01    &  -1.1744e+00 \\
			$\Gamma$ & average of total EGF& -2.1830e-01   &   +3.1164e-01   &   +3.6000e+00   &   +5.0090e+00 \\&&& \\
			
			$\mu_0 $ & EMT factor& -6.6859e-02  &    +9.5767e-02   &   +1.1345e+00   &   +1.5432e+00 \\
			$\mu_{1/2} $ & critical EGF density & +1.7618e-02     & -2.5178e-02    &  -2.8800e-01   &   -4.0878e-01 \\&&& \\
			
			$\mu_D $ & proliferation rate of DCCs&  +2.3304e-01   &   +3.9116e-02  &    +1.2000e-01   &   -6.8597e-04 \\
			$\mu_S $ & proliferation rate of CSCs&   +2.2971e-04   &   +4.9220e-02   &   $\pm$0    &  +4.6599e-03 \\
			$\mu_F $ & proliferation rate of fibroblasts&  -1.3865e-04   &   -5.2592e-05  &    $\pm$0   &   +1.1644e-03 \\
			$\mu_v $ & ECM remodelling rate & +5.2516e-02    &  +1.0917e-02    &  +6.6000e-01    &  +2.4808e-01 \\ &&& \\
			
			$\beta_F $	& apoptosis of fibroblasts &	  +1.2117e-03   &   +4.8903e-04   &   $\pm$0   &   -7.7809e-03 \\
			$\beta_m $ & decay rate of MMPs& -4.7396e-02  &    -9.1276e-03   &   -4.9200e-01    &  -8.6865e-01 \\ & & & \\
			
			$\alpha_D $ & MMP production rate of DCCs & +3.6965e-01   &   +2.6936e-02   &   $\pm$0    &  +7.9042e-01 \\
			$\alpha_S $ & MMP production rate of CSCs & +6.4883e-03   &   +6.6901e-03   &   +6.6000e-01    &  +1.3518e-01 \\&&& \\
			
			$\delta_v $ & ECM degradation rate & -6.8955e-06   &   -4.4457e-06    &  $\pm$0    &  -4.8110e-05 \\ 
			$\mtra $ & transdifferentiation rate & -2.0113e-01   &   -6.7081e-01   &   -1.2000e+01    &  -3.6641e+01 \\

		\end{tabular}\end{center}
		\caption{Sensitivity of model properties with respect to the model parameters in Experiment \ref{exp:1D.main}. A large absolute value in the table indicates a high influence of the parameter (rows) on the corresponding attribute (column). For details on the computation see Section \ref{sec:par.study}, the results are discussed in Section \ref{sec:discussion}.}\label{table:sensitivity}
	\end{table}	
	
	Regarding the sensitivity of the model simulation from  Table \ref{table:sensitivity}, cf. Section \ref{sec:par.study}, we can see that the invasiveness of the CSCs relative to the DCCs (quantified by the front peak distance) is not only affected by the haptotactic sensitivity of both types of cancer cells, but also by the EGF. Indeed an increase in the total amount of EGF, or in the EMT strength has a significant positive impact, while a higher rate of fibroblast transdifferention would weaken the aggressiveness. Further, higher ECM remodelling rates increase the tumor by increasing the mass of both cancer types, the invasiveness of the CS, and also in the density of the invading CSCs (i.e. the CSC front height), see Figure \ref{fig:justify.FIB} (compare bottom and top row). The apoptosis of fibroblasts moreover supports the tumour while having a stronger influence on the DCCs then on the CSCs. In comparison to the transdifferentiation, which weakens the tumour growth overall, the proliferation of the fibroblasts has only a weak negative impact on the tumour size. An increase in the number of EGF receptors on the DCCs would promote the EMT transition and thus the mass, and the invasiveness of the DCCs. On the other hand, increasing the receptors on the CSCs has the inverse but significantly weaker effect.
	
	In Figure \ref{fig:parameter.dependence} we can see the effect that the EMT controlling parameters $\mu_0$, $\Gamma$, $\mu_{1/2}$ have on the mass and on the primary propagating front of the DCCs and the CSCs. In particular we observe that the increase of $\mu_0$ (maximum EMT rate) causes the depletion of the DCCs. 
	Moreover, an increase of $\Gamma$ (average EGF in the environment) increases the amount of CSCs relative to DCCs, until an eventual stagnation of the CSCs aggressiveness, with no change upon further increase in $\Gamma$. Higher values of the constant $\mu_{1/2}$ have a somewhat different effect: the aggressiveness of CSCs (in both density and speed) decays as is their total mass. Hence, increasing $\mu_{1/2}$, or decreasing $\mu_0$ and $\Gamma$ would have a beneficial impact to the patient.
		
	With our parameter investigations we have identified potential treatments that affect the invasiveness and migration of the CSCs. Our model faithfully predicts, that decreasing the total amount of EGF in the environment, or the maximum EMT rate $\mu_0$, or the amount of EGFRs $\ld$, or increasing of  $\mu_{1/2}$,  leads to a direct decrease of the EMT rate $\memt$ as one would expect based on the setting for the model. Furthermore, the dependence of the mass of CSCs and DDCs on the effective EMT-rate agrees with the expectations. But not only the height of the CSC front-peak, but also the difference between the position of the peaks of the two cell populations is positively correlated with the effective EMT-rate and thus with the CSC-mass. Similar results can be obtained by the inhibition of the remodelling of the ECM, or by the enhancement of transdifferentiation, and by the prevention of fibroblast apoptosis.

\begin{figure}[t]
		\resizebox{\linewidth}{!}{
		\begin{tabular}{c | c|c|c}
			&	DCC & CSC & ECM\\
			\rotatebox{90}{\hspace{.8cm}uniform ECM}  &	
			\includegraphics{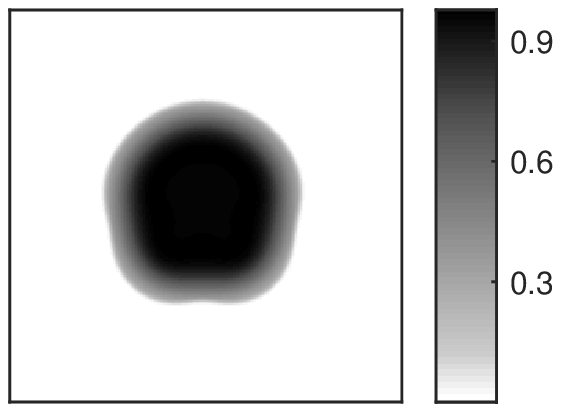} 
			&\includegraphics{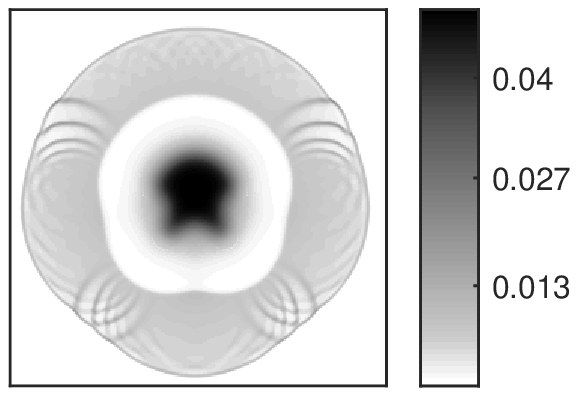}
			&\includegraphics{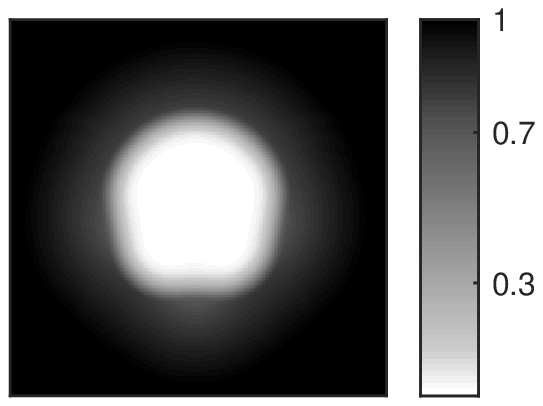} \\
			\hline		
			\rotatebox{90}{\hspace{.3cm}heterogeneous ECM}
			&\includegraphics{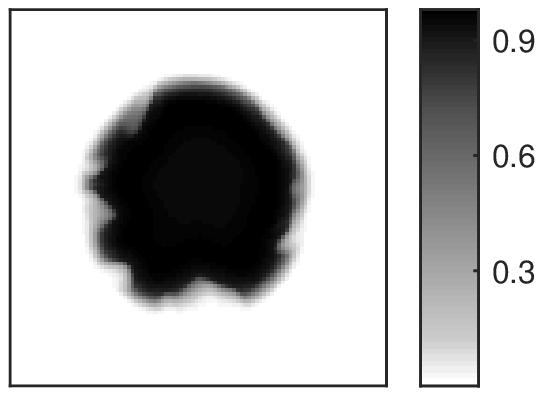} 
			&\includegraphics{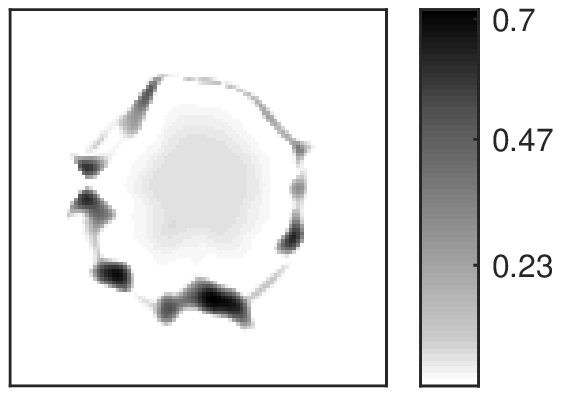}
			&\includegraphics{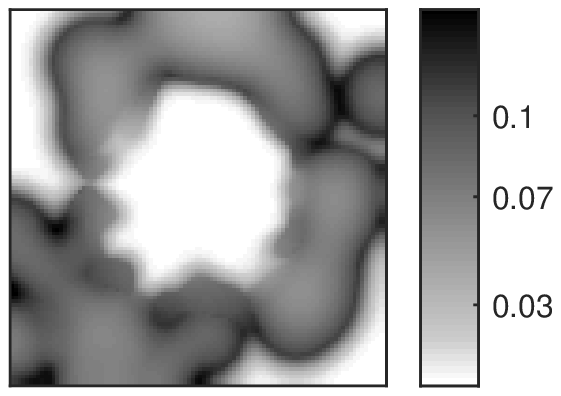}
		\end{tabular}}
		\caption{Comparison of two invasion systems with uniform and non-uniform ECM, respectively. System \eqref{eq:full.system} is resolved with uniform initial ECM distribution (upper row,Experiment \ref{exp:2D.main}), and with non-uniform initial distribution of the ECM (lower row, cf. Experiment \ref{exp:2D.non.uni.ECM.I}). In both cases the ECM is degraded by the MMPs and remodelled by the fibroblast cells. We observe a significant loss of symmetry in the invasion of the CSCs with the higher concentrations following the structure of the ECM.}\label{fig:justify.ECM}	
	\end{figure}

	In Figure \ref{fig:justify.ECM} we can see the effect that an initially non-uniform ECM has on the migration of the cancer. We note in particular, that the invasion is no longer uniform and it is highly influenced by the gradients of the ECM. These can lead to the creation of islands/clusters of CSCs. 
						
\begin{figure}[t]
		\resizebox{\linewidth}{!}{
		\begin{tabular}{c| c|c|c}
			&	DCC & CSC & ECM\\
			\rotatebox{90}{\hspace{1.5cm}$t = 35$}
			&\includegraphics{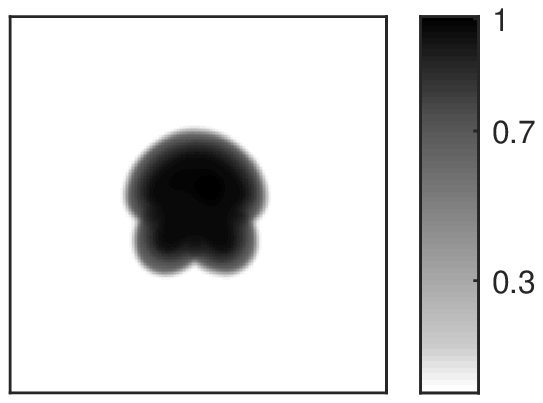} 
			&\includegraphics{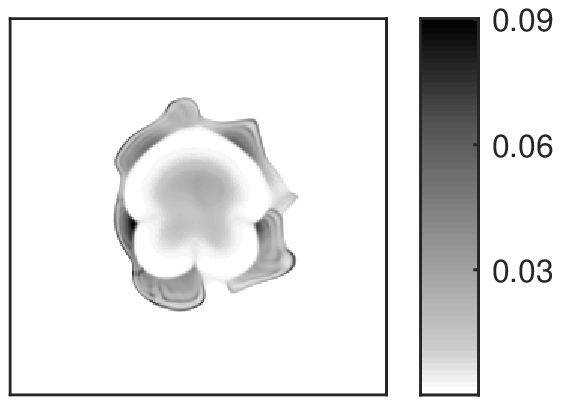}
			&\includegraphics{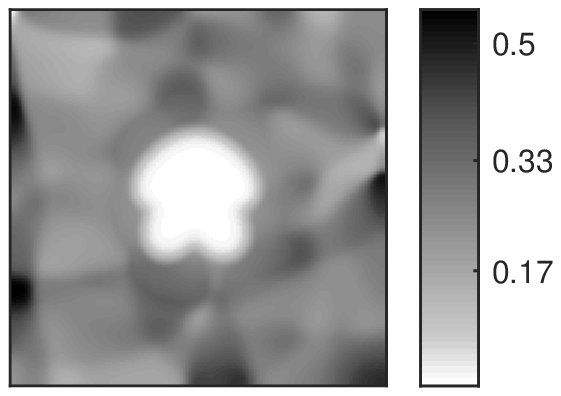} \\
			\hline
			\rotatebox{90}{\hspace{1.5cm}$t = 70$}
			&\includegraphics{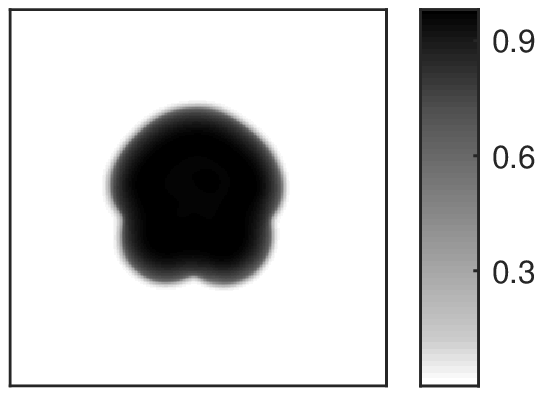} 
			&\includegraphics{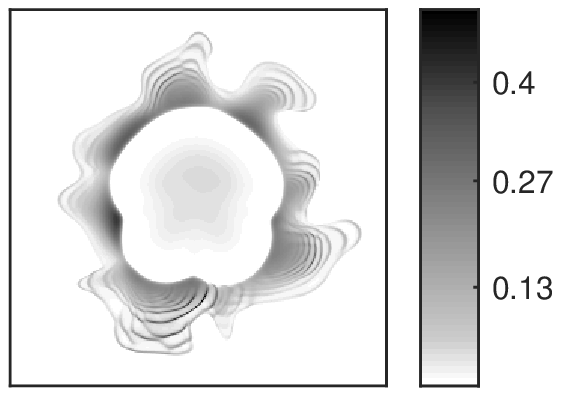}
			&\includegraphics{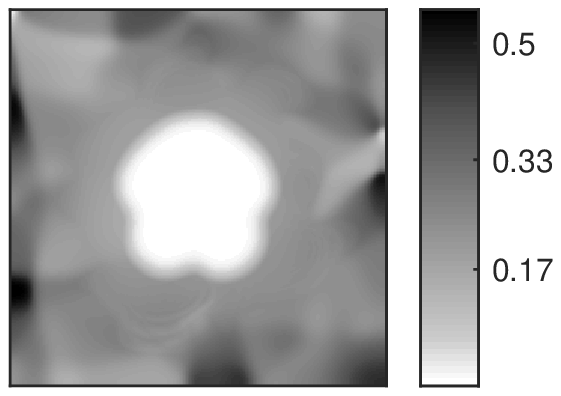} \\
			\hline
			\rotatebox{90}{\hspace{1.5cm}$t = 100$}
			&\includegraphics{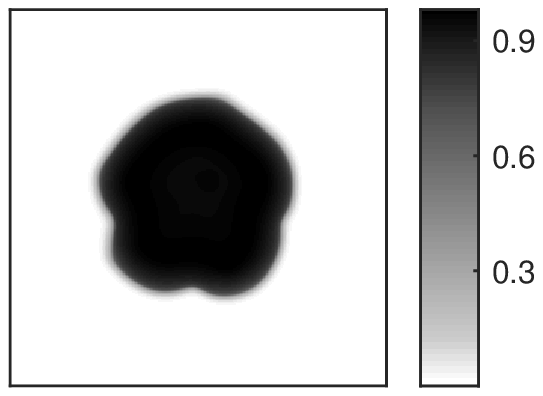} 
			&\includegraphics{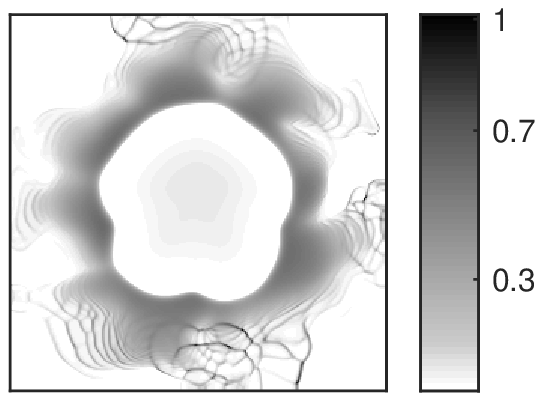}
			&\includegraphics{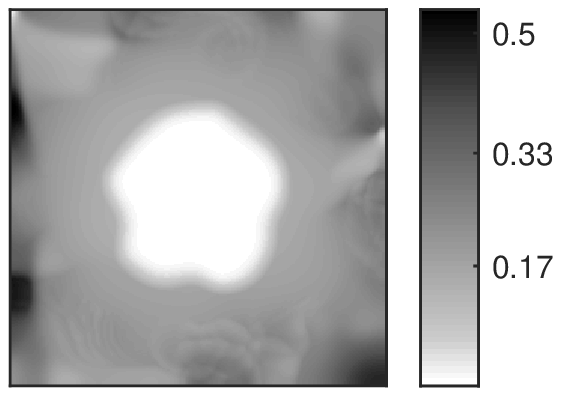}
		\end{tabular}}
		\caption{Invasion of the ECM (right) by the DCCs (left) and CSCs (center) including  all the components of the model: the ECM is assumed to be non-uniform and remodelled by the fibroblast cell, the EMT is EGF-driven, and the proliferation is modelled by the positive-part logistic function. Note the wavy pattern due to the CSCs migration of the ECM,  Experiment \ref{exp:2D.non.uni.ECM.II}.}\label{fig:2D.exp2}	
	\end{figure}

	In Figure \ref{fig:2D.exp2} we present the invasion of the ECM by the cancer cells using the full system \eqref{eq:full.system} and parameters from Table \ref{tbl:params}.  We note that in the invasion of the CSCs nonlinear waves appear, that are created by the interaction of their propagating fronts.
	
\section{Conclusion}

The main goal of this work is to develop a mathematical model and get a deeper insight into the role of EMT in the invasion of cancer. In particular, we integrate the EGF-driven-EMT theory and develop an algebraic-elliptic model \eqref{eq:ellip} to account for the amount of EGF molecules in the vicinity of a tumor. To this end we introduce a non-constant EMT coefficient \eqref{eq:muemt_2} that drives the de-differentiation of DCCs to CSCs. 

This EMT model is combined with a haptotaxis system of the Keller-Segel type modelling the invasion of the ECM by both DCCs and CSCs. We endow the resulting system with a term describing the degradation of the ECM by matrix degenerating MMPs secreted by both families of cancer cells, and by the ECM remodelling driven by cancer associated fibroblast cells \eqref{eq:sub.remod}. The fibroblast cells are assumed to be produced by the CSCs via a cell transdifferentiation program \eqref{eq:sub.trans}. Another particular component of the proposed model is the positive-part logistic proliferation term. Numerically, we resolve the full system in one- and two- spatial dimensions with a robust higher order implicit explicit finite volume scheme developed in our previous works \cite{Sfakianakis.2014b, Sfakianakis.2015}.

With our numerical experiments, we are able to reproduce qualitatively the dynamic behaviour of the DCCs and the CSCs: the CSCs separate from the main body of the tumour, invade the ECM and develop highly dynamic behaviour in the form of merging and emerging concentrations. We also show that during the invasion of the ECM, the CSCs exhibit complex non-linear behaviour in the form of propagating and interacting waves in both cases of initial uniform and non-uniform ECM. 

We are moreover able to deduce information regarding the invasiveness of CSCs. Using the components of the EMT subsystem we can suggest potential scenarios for their control. In particular, performing a sensitivity analysis on the parameters of the models to verify that the parameters and components of the EMT are among the most influential in the invasiveness and aggressiveness of the tumour. With our study, we can see that the CSCs produced by small tumours become detectable only in later stages of the evolution of the cancer; while in the meantime they migrate in a fast pace.


\bibliographystyle{plain} 
\begin{multicols}{2}
	\bibliography{EMT_transition}

\begin{thebibliography}{10}

\bibitem{Alt.1985}
W.~Alt and D.A. Lauffenburger.
\newblock Transient behavior of a chemotaxis system modelling certain types of
  tissue inflammation.
\newblock {\em J. Math. Bio.}, 24(6):691--722, 1987.

\bibitem{Andasari.2011}
V.~Andasari, A.~Gerisch, G.~Lolas, A.P. South, and M.A.J. Chaplain.
\newblock Mathematical modelling of cancer cell invasion of tissue: biological
  insight from mathematical analysis and computational simulation.
\newblock {\em J. Math. Biol.}, 63(1):141--171, 2011.

\bibitem{Anderson.2000}
A.R.A. Anderson, M.A.J. Chaplain, E.L. Newman, R.J.C. Steele, and A.M.
  Thompson.
\newblock Mathematical modelling of tumour invasion and metastasis.
\newblock {\em Comput. Math. Method. M.}, 2(2):129--154, 2000.

\bibitem{Armitage.1954}
P.~Armitage and R.~Doll.
\newblock The age distribution of cancer and a multi-stage theory of
  carcinogenesis.
\newblock {\em Br. J. Cancer}, 8(1):1, 1954.

\bibitem{Bellomo.2008}
N.~Bellomo, N.K. Li, and P.K. Maini.
\newblock On the foundations of cancer modelling: Selected topics,
  speculations, and perspectives.
\newblock {\em Math. Mod. Meth. Appl. S.}, 18(04):593--646, 2008.

\bibitem{Brabletz.2005}
T.~Brabletz, A.~Jung, S.~Spaderna, F.~Hlubek, and T.~Kirchner.
\newblock Opinion: migrating cancer stem cells - an integrated concept of
  malignant tumour progression.
\newblock {\em Nat. Rev. Cancer}, 5(9):744--749, 2005.

\bibitem{bray2001cell}
D.~Bray.
\newblock {\em Cell movements: from molecules to motility}.
\newblock Garland Science, 2001.

\bibitem{Chaplain.2005}
M.A.J. Chaplain and G.~Lolas.
\newblock Mathematical modelling of cancer cell invasion of tissue. the role of
  the urokinase plasminogen activation system.
\newblock {\em Math. Mod. Meth. Appl. S.}, 15(11):1685--1734, 2005.

\bibitem{Cox.2011}
T.R. Cox and J.T. Erler.
\newblock Remodeling and homeostasis of the extracellular matrix: implications
  for fibrotic diseases and cancer.
\newblock {\em Dis. Model. Mech.}, 4(2):165--178, 2011.

\bibitem{Domschke.2014}
P.~Domschke, D.~Trucu, A.~Gerisch, and M.A.J. Chaplain.
\newblock Mathematical modelling of cancer invasion: Implications of cell
  adhesion variability for tumour infiltrative growth patterns.
\newblock {\em J. Theor. Biology}, 361:41--60, 2014.

\bibitem{Eladdadi.2008}
A.~Eladdadi and D.~Isaacson.
\newblock A mathematical model for the effects of {HER}2 overexpression on cell
  proliferation in breast cancer.
\newblock {\em Bull. Math. Biol.}, 70:1707--1729, 2008.

\bibitem{Erler.2009}
J.T. Erler and V.W. Weaver.
\newblock Three-dimensional context regulation of metastasis.
\newblock {\em Clin. Exp. Metastasis}, 26(1):35--49, 2009.

\bibitem{Fan.2013}
Y.L. Fan, M.~Zheng, Y.L. Tang, and X.H. Liang.
\newblock A new perspective of vasculogenic mimicry: E{M}{T} and cancer stem
  cells (review).
\newblock {\em Oncol. Lett.}, 6(5):1174--1180, 2013.

\bibitem{Fisher.1958}
J.C. Fisher.
\newblock Multiple-mutation theory of carcinogenesis.
\newblock {\em Nature}, 181(4609):651--652, 1958.

\bibitem{Ganguly.2006}
R.~Ganguly and I.K. Puri.
\newblock Mathematical model for the cancer stem cell hypothesis.
\newblock {\em Cell Proliferat.}, 39(1):3--14, 2006.

\bibitem{2}
D.~Gao, L.T. Vahdat, S.~Wong, J.C. Chang, and V.~Mittal.
\newblock Microenvironmental regulation of epithelial-mesenchymal transitions
  in cancer.
\newblock {\em Cancer Res.}, 72(19):4883--4889, 2012.

\bibitem{gerisch2006robust}
A.~Gerisch and M.A.J. Chaplain.
\newblock Robust numerical methods for taxis--diffusion--reaction systems:
  {A}pplications to biomedical problems.
\newblock {\em Math. Comput. Model.}, 43(1):49--75, 2006.

\bibitem{Gerisch.2008}
A.~Gerisch and M.A.J. Chaplain.
\newblock Mathematical modelling of cancer cell invasion of tissue: {L}ocal and
  nonlocal models and the effect of adhesion.
\newblock {\em J. Theor. Biol.}, 250(4):684--704, 2008.

\bibitem{Gupta.2009}
P.B. Gupta, C.L. Chaffer, and R.A. Weinberg.
\newblock Cancer stem cells: mirage or reality?
\newblock {\em Nat. Med.}, 15(9):1010--1012, 2009.

\bibitem{Sfakianakis.2015}
N.~Hellmann, N.~Kolbe, and N.~Sfakianakis.
\newblock A mathematical insight in the epithelial-mesenchymal-like transition
  in cancer cells and its effect in the invasion of the extracellular matrix.
\newblock {\em Bull. Braz. Math. Soc.}, 47(1):397--412, 2016.

\bibitem{hundsdorfer2003numerical}
W.~Hundsdorfer and J.G. Verwer.
\newblock {\em Numerical solution of time-dependent
  advection-diffusion-reaction equations}, volume~33.
\newblock Springer, 2003.

\bibitem{imai1982epidermal}
Y.~Imai, C.K.H. Leung, H.G. Friesen, and R.P.C. Shiu.
\newblock Epidermal growth factor receptors and effect of epidermal growth
  factor on growth of human breast cancer cells in long-term tissue culture.
\newblock {\em Cancer Res.}, 42(11):4394--4398, 1982.

\bibitem{jeulin2008egf}
C.~Jeulin, V.~Seltzer, D.~Bailb{\'e}, K.~Andreau, and F.~Marano.
\newblock {E}{G}{F} mediates calcium-activated chloride channel activation in
  the human bronchial epithelial cell line 16{H}{B}{E}14o-: involvement of
  tyrosine kinase p60c-src.
\newblock {\em Am. J. Physiol.-Lung C.}, 295(3):L489--L496, 2008.

\bibitem{Johnston.2010}
M.D. Johnston, P.K. Maini, S.~Jonathan-Chapman, C.M. Edwards, and W.F. Bodmer.
\newblock On the proportion of cancer stem cells in a tumour.
\newblock {\em J. Theor. Biol.}, 266(4):708--711, 2010.

\bibitem{Kaplan.2005}
R.N. Kaplan, R.D. Riba, S.~Zacharoulis, A.H. Bramley, L.~Vincent, C.~Costa,
  D.D. MacDonald, D.K. Jin, K.~Shido, S.A. Kerns, Z.~Zhu, D.~Hicklin, Y.~Wu,
  J.L. Port, N.~Altorki, E.R. Port, D.~Ruggero, S.V. Shmelkov, K.K. Jensen,
  S.~Rafii, and D.~Lyden.
\newblock {VEGFR}1-positive haematopoietic bone marrow progenitors initiate the
  pre-metastatic niche.
\newblock {\em Nature}, 438(7069):820--827, 2005.

\bibitem{Katsumo.2013}
Y.~Katsumo, S.~Lamouille, and R.~Derynck.
\newblock {TGF}-$\beta$~ signaling and epithelial–mesenchymal transition in
  cancer progression.
\newblock {\em Curr. Opin. Oncol.}, 25(1):76--84, 2013.

\bibitem{Kawamoto.1983}
T.~Kawamoto, J.~Mendelsohn, A.~Le, G.H. Sato, C.S. Lazar, and G.N. Gill.
\newblock Relation of epidermal growth factor receptor concentration to growth
  of human epidermoid carcinoma {A}431 cells.
\newblock {\em J. Biol. Chem.}, 259(12):7761--7766, 1984.

\bibitem{KS.1970}
E.F. Keller and L.A. Segel.
\newblock Initiation of slime mold aggregation viewed as an instability.
\newblock {\em J. Theor. Biol.}, 26(3):399--415, 1970.

\bibitem{christopher2001additive}
C.A. Kennedy and M.H. Carpenter.
\newblock Additive {R}unge-{K}utta schemes for convection-diffusion-reaction
  equations.
\newblock {\em Appl. Numer. Math.}, 1(44):139--181, 2003.

\bibitem{Kim.2009}
D.H. Kim, K.~Han, K.~Gupta, K.W. Kwon, K.Y. Suh, and A.~Levchenko.
\newblock Mechanosensitivity of fibroblast cell shape and movement to
  anisotropic substratum topography gradients.
\newblock {\em Biomaterials}, 30(29):5433--5444, 2009.

\bibitem{KF.2010}
Y.~Kim and A.~Friedmann.
\newblock Interaction of tumor with its micro-environment: A mathematical
  model.
\newblock {\em B. Math. Biol.}, 72(5):1029--1068, 2010.

\bibitem{klein2004structure}
P.~Klein, D.~Mattoon, M.A. Lemmon, and J.~Schlessinger.
\newblock A structure-based model for ligand binding and dimerization of {EGF}
  receptors.
\newblock {\em P. Natl. Acad. Sci. USA}, 101(4):929--934, 2004.

\bibitem{Kolbe.2013}
N.~Kolbe.
\newblock Mathematical modelling and numerical simulations of cancer invasion.
\newblock Master's thesis, University of Mainz, supervised by M.
  Luk\'a\v{c}ov\'a-Medvid'ov\'a, and N. Sfakianakis, 2013.

\bibitem{Sfakianakis.2014b}
N.~Kolbe, J.~Kat’uchov{\'a}, N.~Sfakianakis, N.~Hellmann, and
  M.~Luk\'a\v{c}ov\'a-Medvid'ov\'a.
\newblock A study on time discretization and adaptive mesh refinement methods
  for the simulation of cancer invasion : {T}he urokinase model.
\newblock {\em Appl. Math. Comput.}, 273:353--376, 2016.

\bibitem{Lukacova.2012}
A.~Kurganov and M.~Luk\'a\v{c}ov\'a-Medvid'ov\'a.
\newblock Numerical study of two-species chemotaxis models.
\newblock {\em Discrete Cont. Dyn-B}, 19(1):131--152, 2014.

\bibitem{Mani.2008}
S.A. Mani, W.~Guo, M.J. Liao, E.N. Eaton, A.~Ayyanan, A.Y. Zhou, M.~Brooks,
  F.~Reinhard, C.C. Zhang, M.~Shipitsin, L.L. Campbell, K.~Polyak, C.~Brisken,
  J.~Yang, and R.A. Weinberg.
\newblock The epithelial-mesenchymal transition generates cells with properties
  of stem cells.
\newblock {\em Cell}, 133(4):704--715, 2008.

\bibitem{chaplain5}
S.R. McDougall, A.R.A. Anderson, M.A.J. Chaplain, and J.A. Sheratt.
\newblock Mathematical modelling of flow through vascular networks:
  Implications for tumour-induced angiogenesis and chemotherapy strategies.
\newblock {\em Bull. Math. Biol.}, 64(4):673--702, 2002.

\bibitem{Michor.2008}
F.~Michor.
\newblock Mathematical models of cancer stem cells.
\newblock {\em J. Clin. Oncol.}, 26(17):2854--2861, 2008.

\bibitem{7}
A.~Neagu, V.~Mironov, I.~Kosztin, B.~Barz, M.~Neagu, R.A. Moreno-Rodriguez,
  R.R. Markwald, and G.~Forgacs.
\newblock Computational modelling of epithelial–mesenchymal transformations.
\newblock {\em Biosystems}, 100(1):23--30, 2010.

\bibitem{Nordling.1953}
C.O. Nordling.
\newblock A new theory on the cancer-inducing mechanism.
\newblock {\em Br. J. Cancer}, 7(1):68, 1953.

\bibitem{ozcan2006nature}
F.~{\"O}zcan, P.~Klein, M.A. Lemmon, I.~Lax, and J.~Schlessinger.
\newblock On the nature of low-and high-affinity {EGF} receptors on living
  cells.
\newblock {\em P. Natl. Acad. Sci. USA}, 103(15):5735--5740, 2006.

\bibitem{pareschi2005implicit}
L.~Pareschi and G.~Russo.
\newblock Implicit-explicit {R}unge-{K}utta schemes and applications to
  hyperbolic systems with relaxation.
\newblock {\em J. Sci. Comput.}, 25(1-2):129--155, 2005.

\bibitem{Perumpanani.1996}
A.J. Perumpanani, J.A. Sherratt, J.~Norbury, and H.M. Byrne.
\newblock Biological inferences from a mathematical model for malignant
  invasion.
\newblock {\em Invas. Metast.}, 16(4-5):209--221, 1996.

\bibitem{Preziosi.2003}
L.~Preziosi.
\newblock {\em Cancer modelling and simulation}.
\newblock CRC Press, 2003.

\bibitem{Radisky.2005}
D.C. Radisky.
\newblock Epithelial-mesenchymal transition.
\newblock {\em J. Cell Sci.}, 118(19):4325--4326, 2005.

\bibitem{Reya.2001}
T.~Reya, S.J. Morrison, M.F. Clarke, and I.L. Weissman.
\newblock Stem cells, cancer, and cancer stem cells.
\newblock {\em Nature}, 414(6859):105--111, 2001.

\bibitem{robbins1965further}
K.C. Robbins, L.~Summaria, D.~Elwyn, and G.H. Barlow.
\newblock Further studies on the purification and characterization of human
  plasminogen and plasmin.
\newblock {\em J. Biol. Chem.}, 240(1):541--550, 1965.

\bibitem{Shien}
K.~Shien, S.~Toyooka, H.~Yamamoto, J.~Soh, M.~Jida, K.L. Thu, S.~Hashida,
  Y.~Makl, E.~Ichlhara, H.~Asano, K.~Tsukuda, N.~Taglkawa, K.~Klura, A.F.
  Gazdar, W.L. Lam, and S.~Miyoshi1.
\newblock Acquired resistance to {EGFR} inhibitors is associated with a
  manifestation of stem cell–like properties in cancer cells.
\newblock {\em Cancer Res.}, 73(10):3051--3061, 2013.

\bibitem{Czochra.2012}
T.~Stiehl and A.~Marciniak-Czochra.
\newblock Mathematical modeling of leukemogenesis and cancer stem cell
  dynamics.
\newblock {\em Math. Model Nat. Phenom.}, 7(01):166--202, 2012.

\bibitem{Stinner.2015}
C.~Stinner, C.~Surulescu, and A.~Uatay.
\newblock Global existence for a go-or-grow multiscale model for tumor invasion
  with therapy.
\newblock Preprint on webpage at
  http://nbn-resolving.de/urn/resolver.pl?urn:nbn:de:hbz:386-kluedo-42943,
  2015.

\bibitem{stokes1991analysis}
C.L. Stokes and D.A. Lauffenburger.
\newblock Analysis of the roles of microvessel endothelial cell random motility
  and chemotaxis in angiogenesis.
\newblock {\em J. Theor. Biol.}, 152(3):377--403, 1991.

\bibitem{Stokes419}
C.L. Stokes, D.A. Lauffenburger, and S.K. Williams.
\newblock Migration of individual microvessel endothelial cells: stochastic
  model and parameter measurement.
\newblock {\em J. Cell Sci.}, 99(2):419--430, 1991.

\bibitem{Thiery.2002}
J.P. Thiery.
\newblock Epithelial--mesenchymal transitions in tumour progression.
\newblock {\em Nat. Rev. Cancer}, 2(6):442--454, 2002.

\bibitem{Vainstein.2012}
V.~Vainstein, O.U. Kirnasovsky, Y.~Kogan, and Z.~Agur.
\newblock Strategies for cancer stem cell elimination: Insights from
  mathematical modelling.
\newblock {\em J. Theor. Biology}, 298:32--41, 2012.

\bibitem{van1977towards}
B.~Van~Leer.
\newblock Towards the ultimate conservative difference scheme. {IV}. {A} new
  approach to numerical convection.
\newblock {\em J. Comput. Phys.}, 23(3):276--299, 1977.

\bibitem{3}
D.C. Voon, H.~Wang, J.K.W. Koo, J.H. Chai, Y.T. Hor, T.Z. Tan, Y.S. Chu,
  S.~Mori, and Y.~Ito.
\newblock {EMT}-induced stemness and tumorigenicity are fueled by the
  {EGFR}/{R}as pathway.
\newblock {\em PLoS One}, 8(8):e70427, 2013.

\bibitem{Zhu.2011}
X.~Zhu, X.~Zhou, M.T. Lewis, L.~Xia, and S.~Wong.
\newblock Cancer stem cell, niche and {EGFR} decide tumor development and
  treatment response: A bio-computational simulation study.
\newblock {\em J. Theor. Biol.}, 269(1):138--149, 2011.

\end{thebibliography}
\end{multicols}
\newpage 
\appendix
\section*{Appendix}
\section*{Experiment description}\label{sec:exp.desc}
Here we give technical details on the experiments that have been presented in this work. Our one-dimensional simulations have been computed on the domain $\Omega =[0,7.5]$ with the initial conditions \eqref{eq:ic.1d}, and the experiments in two dimensions have been conducted on $\Omega=[-5,5]^2$ with the initial data \eqref{eq:ic.2d.dcc}, \eqref{eq:ic.2d.rest}. Wherever not otherwise stated, we have employed the parameters from Table \ref{tbl:params}.\\

\begin{experiment}{One-dimensional, EGF driven EMT}\label{exp:1D.main}
	With parameters from Table \ref{tbl:params} except for
	$\mu_0 = 0.55, \quad k_D= 0.5,\quad \Gamma = 7.5^{-1}.$
	See also Figure \ref{fig:invasion1D} and \ref{fig:parameter.dependence}. The convergence study (see Table \ref{tbl:conv}) and the sensitivity analysis, see Table \ref{table:sensitivity}, have been conducted on a smaller domain $\Omega_S=[0,3]$.
\end{experiment}

\vskip 1 em
\begin{experiment}{One-dimensional, adjusted EMT}\label{exp:1D.var.EMT}
	The parameters have been adjusted, so that the invasiveness of the CSCs coincides with the Experiment \ref{exp:1D.ctEMT}:
	$\mu_0 = 0.034,\quad \mu_{1/2}= 0.4,\quad k_D= 0.2, \quad \lambda^D = 3,\quad \Gamma = 7.5^{-1}.$
	See also Figure \ref{fig:compEMT} (middle row).
\end{experiment}

\vskip 1 em
\begin{experiment}{One-dimensional, adjusted EMT \& low EGF}\label{exp:1D.var.EMT.low.EGF}
	Same as Experiment \ref{exp:1D.var.EMT}, with low EGF concentration: $\mu_0 = 0.034,\quad \mu_{1/2}= 0.4,\quad k_D= 0.2, \quad \lambda^D = 3,\quad \Gamma = 10^{-3}.$ See also Figure \ref{fig:compEMT} (lower row).
\end{experiment}

\vskip 1 em
\begin{experiment}{Adjusted EMT \& low initial DCC concentration}\label{exp:1D.var.EMT.small.IC}
	Same parameters as in Experiment \ref{exp:1D.var.EMT}. Advection and diffusion terms were neglected. The initial conditions 
	$c_0^D = 10^{-3},\quad c_0^S = 0, \quad c_0^F= 0,\quad v_0= 1,\quad m_0 = 10^{-6}$ have  been employed.
	See also Figure \ref{fig:compEMTODE} (left).
\end{experiment}

\vskip 1 em
\begin{experiment}{One-dimensional, constant EMT rate}\label{exp:1D.ctEMT}
Our EMT model from Section \ref{sec:EMT.mod} has been neglected, and the EMT transition coefficient in system \eqref{eq:full.system} has been taken as a constant: $\memt = 0.017$. The invasiveness of the CSCs coincides with Experiment \ref{exp:1D.var.EMT}. See Figure \ref{fig:compEMT} (upper row).
\end{experiment}

\vskip 1 em
\begin{experiment}{Constant EMT rate \& low initial DCC concentration}\label{exp:1D.ct.EMT.small.IC}
	Same parameters as in Experiment \ref{exp:1D.ctEMT}. Advection and diffusion terms were neglected. The initial conditions 
	$c_0^D = 10^{-3},\quad c_0^S = 0, \quad c_0^F= 0,\quad v_0= 1,\quad m_0 = 10^{-6}, $ have been employed as in Experiment \ref{exp:1D.var.EMT.small.IC}. See also Figure \ref{fig:compEMTODE} (right).
\end{experiment}

\vskip 1 em
\begin{experiment}{Two-dimensional, uniform ECM \& fibroblast remodelling}\label{exp:2D.main} Parameters from Table \ref{tbl:params}.
		See Figures \ref{fig:invasion2D} and \ref{fig:justify.FIB} (lower row). 
\end{experiment}

\vskip 1 em
\begin{experiment}{Two-dimensional, uniform ECM \& no remodelling}\label{exp:2D.no.remod}
	Same as Experiment \ref{exp:2D.main}, without matrix remodelling: $\mu_v = 0$. See Figure \ref{fig:justify.FIB} (upper row).
\end{experiment}

\vskip 1 em
\begin{experiment}{Two-dimensional, uniform ECM \& self remodelling}\label{exp:2D.self.remod}
	Same as Experiment \ref{exp:2D.main},  with self-remodelling, i.e. the term $+ \mu_v\, \cf\, (1-\cd-\cc-\cf-v)^+$ for the ECM in system \eqref{eq:full.system} has been replaced by $+ \mu_v\, v\, (1-\cd-\cc-\cf-v)^+$, hence the fibroblasts have been neglected. See Figure \ref{fig:justify.FIB} (middle row).
\end{experiment}

\vskip 1 em
\begin{experiment}{Two-dimensional, uniform ECM \& fibroblast remodelling \& smooth initial data}\label{exp:2D.conv}Same as Experiment \ref{exp:2D.main}
	with smooth initial data for the convergence study, see Table \ref{tbl:conv}:
	$$	\cd_0(\mathbf x) =  \exp(-\|\mathbf x\|_2^2), \ \cc_0(\mathbf x) = \cf_0(\mathbf x)=0,
	\ v_0(\mathbf x) = 1 - \cd_0(\mathbf x), \ m_0(\mathbf x) = \frac{1}{20} \, \cd_0(\mathbf x), \quad \mathbf x=(x_1,x_2)\in \Omega\,.$$
\end{experiment}

\vskip 1 em
\begin{experiment}{Two-dimensional, non-uniform ECM }\label{exp:2D.non.uni.ECM.I}
	Same as Experiment \ref{exp:2D.main} with an ECM initial concentration $v_0$ different from \eqref{eq:ic.2d.rest}. 
	The non-uniform initial ECM $v_0$ has been constructed by a random distribution of fibroblast cell concentrations acting without other influences on the domain. We have conducted the simulation on a 100 $\times$ 100 computational grid (lower resolution as opposed to the other experiments). See Figure \ref{fig:justify.ECM} lower row.
\end{experiment}

\vskip 1 em
\begin{experiment}{Two-dimensional, non-uniform ECM }\label{exp:2D.non.uni.ECM.II}
	Similar to Experiment \ref{exp:2D.non.uni.ECM.I}, the non-uniform initial ECM $v_0$ has been constructed by a random distribution of fibroblast cell concentrations acting on the domain. As a result the initial ECM concentration is different from the case in Experiment \ref{exp:2D.non.uni.ECM.I}. We have conducted the simulation using the standard grid resolution with 250 $\times$ 250 mesh cells. See also Figure \ref{fig:2D.exp2}.
\end{experiment}

\end{document}